\documentclass[aps,pre,twocolumn,showpacs,amsfonts,amssymb,floatfix]{revtex4-1}
\usepackage[T1]{fontenc}
\usepackage[utf8]{inputenc}
\usepackage{float}
\usepackage{graphicx}
\usepackage{amssymb}
\usepackage{amsmath}
\usepackage{enumerate}
\usepackage{tikz}
\usetikzlibrary{decorations.pathreplacing}

\makeatletter

\newcommand*{\diff}{\mathop{\!\mathrm{d}\!}}
\date{December 16, 2015} 

\makeatother

\begin{document}

\title{Random Recursive Trees and the Elephant Random Walk}

\author{Rüdiger Kürsten}
\affiliation{Institut für Theoretische Physik, Universität Leipzig, POB 100 920, D-04009 Leipzig, Germany}
\affiliation{International Max Planck Research School Mathematics in the Sciences, Inselstraße 22, D-04103 Leipzig, Germany}

\begin{abstract}
	One class of random walks with infinite memory, so called elephant random walks, are simple models describing anomalous diffusion. We present a surprising connection between these models and bond percolation on random recursive trees. We use a coupling between the two models to translate results from elephant random walks to the percolation process. We calculate, besides other quantities, exact expressions for the first and the second moment of the root cluster size and of the number of nodes in child clusters of the first generation. We further introduce a new model, the skew elephant random walk and calculate the first and second moment of this process.
\end{abstract}
\maketitle

\section{Introduction\label{sec:1}}

Anomalous Diffusion appears in many physical, biological or social systems. Examples are
the traveling motion of persons or money \cite{BHG06},
macro-molecules in cytoplasma \cite{WEKT04},
the motion on membranes \cite{SSS97},
diffusion in porous materials \cite{KB88} or
diffusion in polymer networks \cite{anomalactin04} to name only a few.
Theoretical models describing such phenomena often include a fractal structure of space \cite{MN68,GAA83} or incorporate memory effects \cite{SM75, MK00, MK04, ST04}.

The elephant random walk (ERW) introduced in \cite{ST04} is one of the simplest models leading to anomalous diffusion. Since its appearance about one decade ago the ERW and several modifications of it, see e.g. \cite{CSV07, KHL10, CVFS12, CVS13, HKL14, Kim14, SVC15}, have been the subject of intense research. As we will demonstrate in this paper there is a surprising connection between ERW-like models and percolation on random recursive trees (RRTs) that seems to be unknown in the literature.

RRTs appear naturally in Yule processes \cite{Yule25}. There, an individual reproduces with a constant rate. When the child is born, it evolves independently from its ancestor and they both reproduce with the same rate. The resulting genealogical tree is a random recursive tree. If the possibility of mutations is taken into account in a simple way, percolation clusters of random recursive trees are obtained naturally as genealogical trees, assuming that with some fixed probability, a child forms a new species due to mutation.

Versions of Yule processes have been used to study the distribution of words in prose samples, the distribution of scientists by the number of papers, the distribution of cities by population or the distribution of income \cite{Simon55} and they share certain features with scientific citation networks \cite{Price65} and preferential attachment models \cite{BA99}.

In recent years there is active research on bond percolation on random recursive trees from a mathematical point of view, see e.g. \cite{Bertoin14, Baur14, BB14}. There have been also studies that deal with the cutting of RRTs \cite{MM74, IM07, KP14} and with fires on RRTs \cite{Marzouk14}. Also fragmentation processes induced by deletion of edges \cite{BB15} or removal of vertexes \cite{KB15} are closely related to bond or site percolation on RRTs.

In this paper we use another formulation of the ERW model that was first investigated in Ref. \cite{Kim14} without mentioning the one-to-one equivalence \cite{Kuersten15} to the original model \cite{ST04}. Furthermore we develop and analyze a generalization of the ERW, the skew elephant random walk (SERW), and we employ another generalization of the ERW that was introduced and analyzed in \cite{KHL10}. The ERW and alike models are understood quite well and we can translate many results into the framework of percolation on random recursive trees.

The paper is organized as follows. In Sec.~\ref{sec:2} we introduce the ERW model and discuss two ways to reformulate it. In Sec.~\ref{sec:3} we introduce the memory tree and memory forest of the ERW as a stochastic coupling between the ERW and percolation on RRTs. In Sec.~\ref{sec:4} we discuss in detail the Bernoulli bond percolation on RRTs and calculate exact expressions for expectation values of the root cluster size, the root cluster size squared, of the number of nodes in child-clusters of the $k$th generation, of the number of nodes in first generation child clusters squared, the probability of the root to be isolated and the expectation value of the  number of clusters of size one, using results from ERW, the newly introduced skew elephant random walk (SERW) or the process introduced in \cite{KHL10}.
In Sec.~\ref{sec:6} we use the results of Sec.~\ref{sec:4} to obtain the root cluster size in certain limiting cases that have been studied in the literature. We prove mean square convergence in contrast to the weaker convergence in probability, proved in \cite{Bertoin14,Baur14}, which makes our findings on the root cluster size stronger than those previous results.
In Sec.~\ref{sec:7} we discuss how our findings on percolation on RRTs lead to a deeper understanding of ERW-like models. The model introduced in \cite{KHL10} can be reduced to ERW with a random effective time, which is the root cluster size of a percolation process on a RRT. The appearance of superdiffusion in ERW can be explained by a single huge memory component of ERW, which is also equivalent to a root cluster of the percolation process.

\newpage
\section{The Elephant Random Walk\label{sec:2}}
In order to make this paper self-contained we will review the main results of Schütz and Trimper \cite{ST04} in this section.
They introduced the elephant random walk (ERW). We will also mention the formulation of Kim \cite{Kim14}. He reformulated the ERW model for some parameter regime and reobtained the same results as \cite{ST04} with the same methods but he was not mentioning the one-to-one correspondence between the two models. We will also introduce a similar reformulation as that used in \cite{Kim14}, covering the other parameter regime.

In \cite{ST04} the ERW is introduced as follows. The elephant starts at time $t=0$ and position $x_0=0$ and moves in discrete time on the integers. We denote the position of the elephant at time $t$ by $x_t$. In each time step the elephant moves a distance $\sigma_{t}$. Hence, the elephants position satisfies
\begin{align}
	x_{t+1} = x_{t} + \sigma_{t+1},
	\label{eq:timeevolution}
\end{align}
where $\sigma_{t}$ are random variables, which can take one of the values $\pm 1$. 

In the first step the elephant may move right to $x_{1}=+1$, that is $\sigma_1=+1$ with probability $q\in [0,1]$ or it moves left to $x_{1}=-1$ that is $\sigma_1=-1$ with probability $1-q$.

At any later time $t$ a random number $t'$ is chosen from $\{1, \dots, t-1 \}$ with uniform probability $1/(t-1)$. Then 
\begin{align}
	\sigma_{t+1} = \begin{cases} \phantom{-}\sigma_{t'} \qquad \text{with probability }\tilde{p},\\
		-\sigma_{t'} \qquad \text{with probability }1-\tilde{p}. \end{cases}
	\label{eq:increment}
\end{align}
That means the elephant chooses some random time $t'$ from the past to remember. With probability $\tilde{p}$ it is doing the same as in the past and with probability $1-\tilde{p}$ it is doing the opposite.

Using the abbreviations
\begin{align}
	\alpha = 2 \tilde{p}-1,\notag \\
	\beta = 2 q -1,
	\label{eq:abbreviations}
\end{align}
the first and the second moment have been calculated in \cite{ST04} resulting in
\begin{align}
	\langle x_{t}\rangle &= \beta \frac{\Gamma (\alpha+t)}{\Gamma (t) \Gamma (\alpha+1) } \text{ for }t>1,
	\label{eq:firstmoment}
	\\
	\langle x_{t}^{2}\rangle &= \frac{t}{2\alpha -1}  \left[ \frac{\Gamma (t+2\alpha)}{\Gamma(t+1)\Gamma(2\alpha)} -1 \right].
	\label{eq:secondmoment}
\end{align}
Consider $\beta=0$ for simplicity. Then the first moment is zero $\langle x_t\rangle =0$ and for large times the mean squared displacement is approximately \cite{ST04}
\begin{align}
	\langle x_{t}^{2}\rangle \approx \begin{cases} \frac{t}{3-4\tilde{p} } &\text{for } \tilde{p}<3/4 \\
		t \ln t &\text{for } \tilde{p}=3/4 \\
		\frac{t^{4\tilde{p}-2}}{(4\tilde{p}-3)\Gamma(4\tilde{p}-2)} &\text{for } \tilde{p}>3/4. \end{cases}
	\label{eq:meansquareddisplacement}
\end{align}
This means the system shows normal diffusion for $\tilde{p}<3/4$ and superdiffusion for $\tilde{p}>3/4$.
Therefore the ERW is interesting from a theoretician's point of view as it is one of the simplest models that show anomalous diffusion.

For $\tilde{p}> 1/2$ the model can be reformulated as follows.
In each time step starting from the second the elephant remembers some point in the past.
Then it needs to decide whether to do the same or the opposite as it did in the past.

For that purpose we can chose a uniform random number $r$ between zero and one.
If $r\le \tilde{p}$ the elephant does the same as it did in the past and if $r> \tilde{p}$ it does the opposite.
We will use the notion that the elephant flips a coin that shows heads in the first case and tails in the second.

Now imagine we choose some parameter $p$, such that $0<p< \tilde{p}$ and we modify the coin, such that it shows heads if $r\le p$ and tails otherwise.

In case the coin shows heads it is clear that $r\le p < \tilde{p}$ and the elephant has to do the same as in the past. If on the other hand the coin shows tails the elephant does not know what to do since although $r>p$ non of the possibilities $r\le \tilde{p}$ and $r> \tilde{p}$ can be ruled out.

In that case the elephant needs a second coin.
Assume that if $p<r\le \tilde{p}$ the second coin will show heads and if $r> \tilde{p}$ it will show tails.
Hence the elephant will do the same as in the past when the second coin shows heads and it will do the opposite when the second coin shows tails.
Then the total probabilities of doing the same or opposite as in the past are still the same
\begin{align}
	Pr(\sigma_{t}=\sigma_{t'})&= Pr(\text{1st coin=heads}) \notag\\
	&+ Pr(\text{1st coin=tails})Pr(\text{2nd coin=heads})\notag \\
	&=p + (1-p)\frac{\tilde{p}-p}{1-p}=\tilde{p},\\
	Pr(\sigma_{t}=-\sigma_{t'})&=Pr(\text{1st coin=tails})Pr(\text{2nd coin=tails})\notag\\
	&=(1-p)\frac{1-\tilde{p}}{1-p}=1-\tilde{p}.
	\label{eq:probdiffsame}
\end{align}

We assumed $\tilde{p}>1/2$, therefore $2\tilde{p}-1$ is between zero and one. Hence we are free to choose
\begin{align}
	p=2\tilde{p}-1.
	\label{eq:parameterrelation}
\end{align}
In this case the second coin shows heads and tails with equal probability $1/2$. Thus once the first coin shows tails, the elephant will do the same or the opposite as it has done in the past with equal probability.
From this follows that it will go right or left with equal probability $1/2$ no matter what it has done in the past.
Therefore the elephant can just as well decide to go to the right if the second coin shows heads and to go to the left if the second coin shows tails.

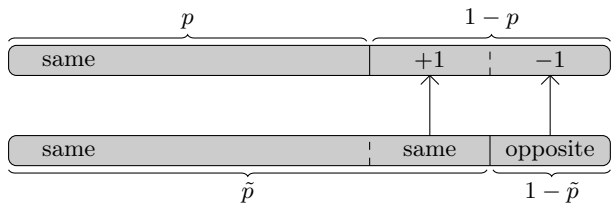
\begin{figure}[b]
\begin{center}
\begin{tikzpicture}[xscale=0.4, yscale=0.4, decoration=brace]
	\filldraw[fill=gray!40, draw=black, rounded corners=3pt] (0,3) rectangle (20,4);
	\filldraw[fill=gray!40, draw=black, rounded corners=3pt] (0,0) rectangle (20,1);
	\draw (12,3) -- (12,4);
	\draw (14,1) -- (14,3);
	\draw (13.7,2.7) -- (14,3);
	\draw (14.3,2.7) -- (14,3);
	\draw (18,1) -- (18,3);
	\draw (17.7,2.7) -- (18,3);
	\draw (18.3,2.7) -- (18,3);
	\draw[decorate, yshift=1ex] (0, 4) -- node[above=0.4ex] {$p$} (11.9, 4);
	\draw[decorate, yshift=1ex] (12.1, 4) -- node[above=0.4ex] {$1-p$}(20, 4);
	\draw[decorate, yshift=-1ex]  (15.9, 0) -- node[below=0.4ex] {$\tilde{p}$}(0, 0) ;
	\draw[decorate, yshift=-1ex]  (20, 0) -- node[below=0.4ex] {$1-\tilde{p}$}(16.1, 0) ;
	\path (2, 3.5) node {same}
	(2, 0.5) node {same}
	(14, 3.5) node {$+1$}
	(18, 3.5) node {$-1$}
	(18, 0.5) node {opposite}
	(14, 0.5) node {same};
	\draw (16,0) -- (16,1);
	\draw[dashed] (16,3) -- (16,4);
	\draw[dashed] (12,0) -- (12,1);
\end{tikzpicture}
\end{center}
\caption{Illustration of the connection between the formulations ERW and ERWv1.
The upper bar represents ERWv1 and the lower bar ERW.
There, with probability $\tilde{p}$ the elephant does the same as in the past and with probability $1-\tilde{p}$ it does the opposite.
The event of doing the same is split into two parts, where the first part has probability $p$ and the second part has probability $\tilde{p} -p=1-\tilde{p}$.
Going to the formulation ERWv1 the first part is kept. The second part and the event of doing the opposite have the same probability, considering only these two events the probability of going right or left is $1/2$ independent on the step that was done in the past. Hence these two events are replaced by the events of going to the right or going to the left in ERWv1, without changing any transition probabilities.\label{fig:reformulation}}
\end{figure}

This is exactly what was done in \cite{Kim14}. There the elephant does the same as it did in the past with probability $p$ and with probability $1-p$ it chooses at random (with equal probability) to go right or left independent on the past. Naturally, the moments obtained in \cite{Kim14} agree with the ones from the original work \cite{ST04} when the parameter identification \eqref{eq:parameterrelation} is taken into account. Confer \cite{Kuersten15} to see exactly how the processes can be defined on the same probability space.
We will refer to this formulation of the ERW as ERWv1. In Fig.~\ref{fig:reformulation} we illustrate the connection between these two formulations.

In case of $\tilde{p}=1/2$ the ERW is just a simple random walk where the elephant goes left or right in each time step with equal probability $1/2$ independent on the past.

When $\tilde{p}< 1/2$ we can use a similar two-coin formulation as ERWv1, but with
\begin{align}
	p=1-2\tilde{p},
	\label{eq:parameterrelation2}
\end{align}
 and the only difference that the elephant will do the opposite as it did in the past when the first coin shows heads.
That is, it will do the opposite as it did in the past with probability $p$ and with probability $1-p$ it will choose at random whether to go right or left. We refer to this formulation as ERWv2.

\section{Memory Tree of the Elephant\label{sec:3}}

Consider the formulation ERWv1.
We are going to draw the memory tree of the elephant.
For each time step we will add a node which will be labeled by the number of the time step and a spin. The spin of the node with number $t$ will be $\sigma_{t}$. Hence it is either plus or minus one depending on the direction the elephant moves at time $t$.
For the spin we might shortly write only $+$ or $-$ instead of $\sigma_{t}=\pm 1$, and we will justify calling them ``spins'' in a moment.

We start drawing a node for the first step.
It will get label $1$ and spin $+$ if $\sigma_{1}=+1$ and spin $-$ if $\sigma_{1}=-1$.
We continue to add the second node.
As the elephant will remember what it did in the first step, that is $t'=1$, we will connect the second with the first node.
However, we will delete this edge if the first coin shows tails.
In this case the elephant does not really remember but it chooses at random what to do.
The second node gets the label $2$ and spin $+$ or $-$ depending on the value of $\sigma_{2}$.

We continue to add nodes for each time step and connect them to the node of the point $t'$ in the past that the elephant is remembering.
And when the first coin shows tails we delete this edge.

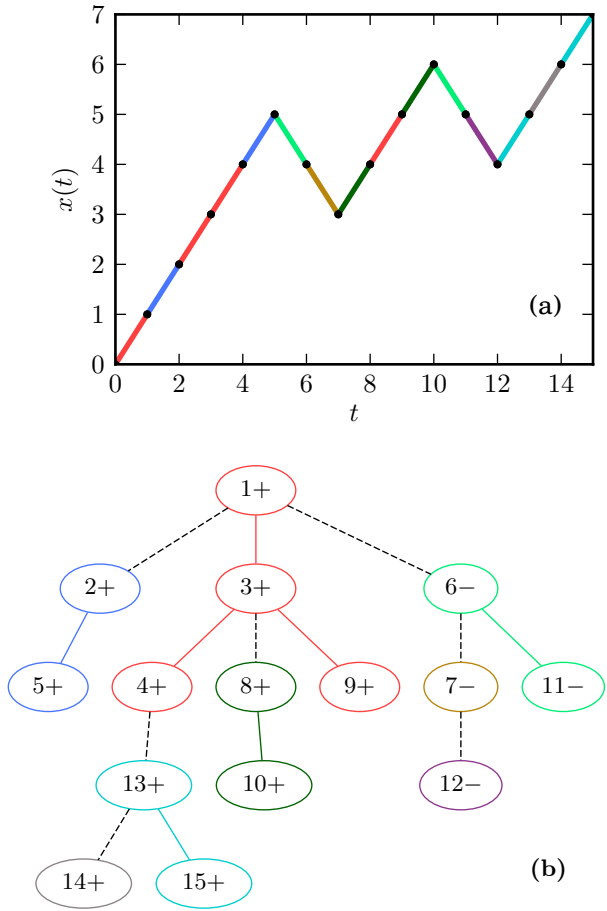
\begin{figure}
	\includegraphics{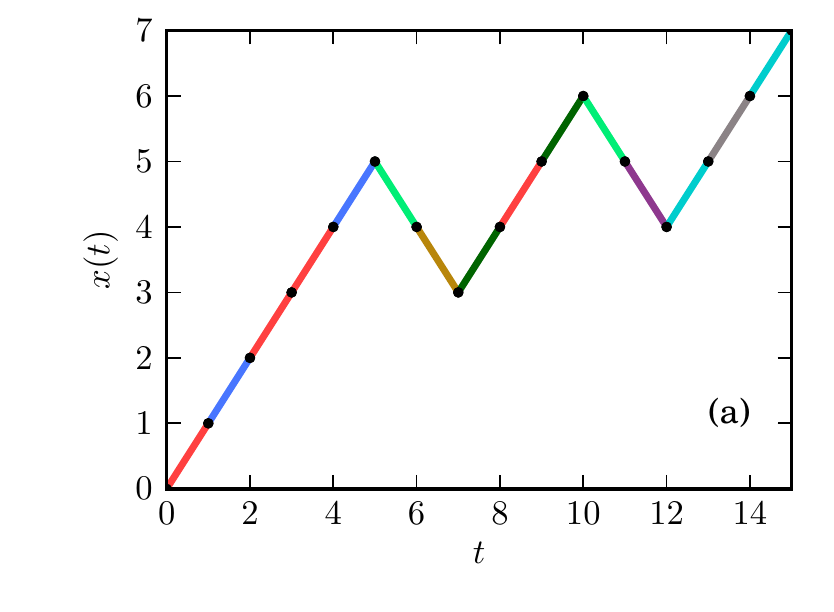}
\def\svgwidth{0.45\textwidth}
\begingroup%
  \makeatletter%
  \providecommand\color[2][]{%
    \renewcommand\color[2][]{}%
  }%
  \providecommand\transparent[1]{%
    \renewcommand\transparent[1]{}%
  }%
  \providecommand\rotatebox[2]{#2}%
  \ifx\svgwidth\undefined%
    \setlength{\unitlength}{444bp}%
    \ifx\svgscale\undefined%
      \relax%
    \else%
      \setlength{\unitlength}{\unitlength * \real{\svgscale}}%
    \fi%
  \else%
    \setlength{\unitlength}{\svgwidth}%
  \fi%
  \global\let\svgwidth\undefined%
  \global\let\svgscale\undefined%
  \makeatother%
  \begin{picture}(1,0.74774775)%
    \put(0,0){\includegraphics[width=\unitlength]{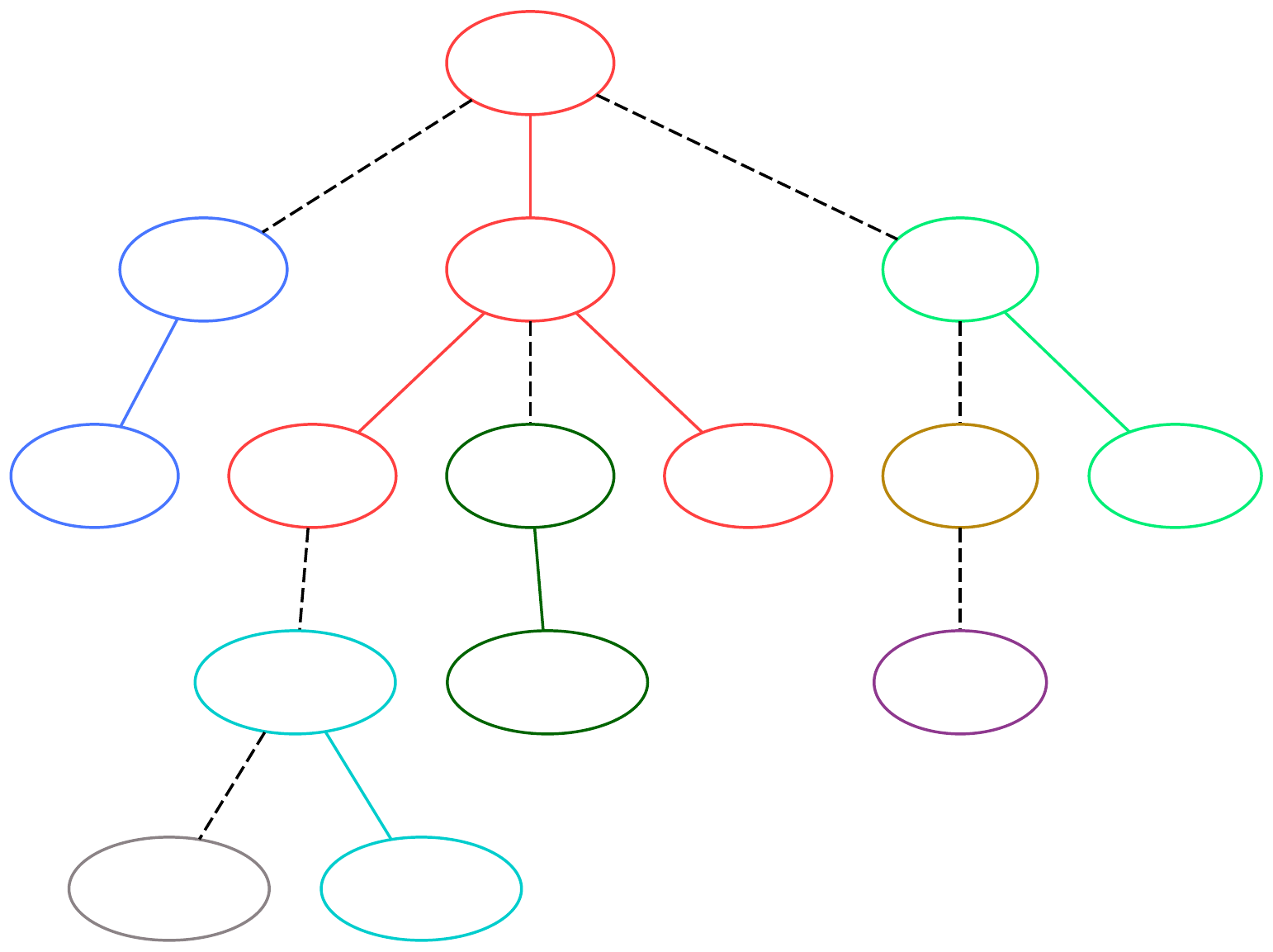}}%
    \put(0.41666667,0.68896395){\makebox(0,0)[b]{\smash{$1 +$}}}%
    \put(0.15990991,0.52680179){\makebox(0,0)[b]{\smash{$2 +$}}}%
    \put(0.41666667,0.52680179){\makebox(0,0)[b]{\smash{$3 +$}}}%
    \put(0.7545045,0.52680179){\makebox(0,0)[b]{\smash{$6 -$}}}%
    \put(0.07432432,0.36463963){\makebox(0,0)[b]{\smash{$5 +$}}}%
    \put(0.2454955,0.36463963){\makebox(0,0)[b]{\smash{$4 +$}}}%
    \put(0.41666667,0.36463963){\makebox(0,0)[b]{\smash{$8 +$}}}%
    \put(0.58783784,0.36463963){\makebox(0,0)[b]{\smash{$9 +$}}}%
    \put(0.23198198,0.20247748){\makebox(0,0)[b]{\smash{$13 +$}}}%
    \put(0.7545045,0.36463963){\makebox(0,0)[b]{\smash{$7 -$}}}%
    \put(0.92342342,0.36463963){\makebox(0,0)[b]{\smash{$11 -$}}}%
    \put(0.7545045,0.20247748){\makebox(0,0)[b]{\smash{$12 -$}}}%
    \put(0.43018018,0.20247748){\makebox(0,0)[b]{\smash{$10 +$}}}%
    \put(0.13288288,0.04031531){\makebox(0,0)[b]{\smash{$14 +$}}}%
    \put(0.33108108,0.04031531){\makebox(0,0)[b]{\smash{$15 +$}}}%
    \put(0.868,0.05237899){\color[rgb]{0,0,0}\makebox(0,0)[lb]{\textbf{(b)}}}%
  \end{picture}%
\endgroup%
	\caption{(color online) Trajectory (a) and memory forest (b) of the same realization of ERWv1. Deleted edges are indicated by dashed lines. Time steps, or nodes, that belong to one memory component, that is one cluster, are drawn with the same color. $q=1/2$, $p=0.7$.\label{fig:erwv1}}
\end{figure}

When the elephant made $N$ steps we end up with a graph with $N$ nodes.
If we wouldn't have deleted edges the graph would be a tree. With some missing edges it is a forest.
We call them the memory tree and the memory forest of the elephant, respectively. In Fig.~\ref{fig:erwv1} we show one realization of the ERWv1 until time $t=15$ together with the corresponding memory forest.

There are some interesting observations apparent from the graph.
Following some path of the memory tree starting from the root (the node with label $1$), labels are always increasing along the path.
That is because we attached nodes with higher labels to previously existing ones with lower labels.
The second important observation is that all nodes from one connected component of the memory forest, which we also call a cluster, carry the same spin.
This has the simple reason that each time an edge was not removed, that is the first coin showed heads, the elephant did the same as in the past.

At this stage it is interesting to review how randomness comes into the construction of the graph.
It enters in three stages.
The first stage is the construction of the memory tree.
There it is chosen at random to which node the new one should be attached, that is, which time point in the past the elephant chooses to remember.

The next stage is the deletion of edges, that is the tossing of the first coin.
The third stage is the flipping of the second coin.

In fact these three random influences are independent of each other.
For instance, it will not affect the result of the first coin-flip, which point in the past the elephant has chosen to remember.
True is, that there is no need to flip the second coin if the first one shows heads.
However, it can be flipped anyway without influencing the resulting graph.
By the choice of the memory tree and the tossing of the first coin, a forest is constructed and by the flipping of the second coin a spin is assigned to each cluster.
Important is that the construction of the forest and the assignment of spins are really independent.

We observe furthermore that the position of the elephant after $N$ steps is just the sum of the spins of all nodes.
Hence it corresponds to some kind of ``magnetization'' of the graph, which is why we used the notion of spins in the first place. Denoting this magnetization of the memory forest of size $N$ by $M_{N}$ we have
\begin{align}
	x_{t=N}=M_{N}=\sum_{i=1}^{N} \sigma_{i}.
	\label{eq:magnetization}
\end{align}

\section{Deleting Edges of Random Recursive Trees\label{sec:4}}

Recursive trees are non-planar trees.
That means the order of the child nodes of one node does not matter.
Furthermore a recursive tree of size $N$ is labeled with numbers from one to $N$, where the node labeled with $1$ is called the root of the tree.
As already mentioned in the previous section a recursive tree is defined by the property that all paths starting from the root follow nodes with increasing labels.

We can obtain a random recursive tree (RRT) of size $N$ in two ways.
The first is to choose one of all different recursive trees of size $N$ at random with uniform probability.
This is more a definition than a practical way to obtain a random recursive tree.
The second procedure is constructive and recursive. In the first step start with the root node.
In the following $N-1$ steps attach the next node (with next highest label) to one of the existing nodes chosen at random with uniform probability.

RRTs have interesting fractal or self-similar properties.
If one edge of a RRT is deleted, two independent RRTs emerge.
More formal, if we choose a RRT of size $N$ and delete a randomly chosen edge, we obtain a tree that contains the root and another tree.
Conditioned on the size of the root-containing tree to be $k<N$, the root-containing tree is a RRT of size $k$ and the other tree is an independent RRT of size $N-k$ \cite{MM74}, cf. also \cite{BB14}.

In percolation theory one investigates usually the clusters of a random graph that is obtained from another graph by deleting either randomly chosen edges (bond percolation) or vertexes (site percolation) \cite{Grimmett99}.
Often the original graph is a lattice of dimension $d$, e.g. $\mathbb{Z}^{d}$.
Some authors investigated the deletion of vertexes \cite{KB15} or edges \cite{Bertoin14, Baur14, BB14, Marzouk14, BB15} of RRTs.
If each edge is deleted independently with probability $1-p$ or kept with probability $p\in[0,1]$, the process is called Bernoulli bond percolation on RRTs.
This is exactly what we did to obtain the memory forest of the elephant.
We can use the correspondence between the ERW and percolation on RRTs to translate results from one model to the other and vice versa.

We denote the clusters of the memory forest by $c_i$, where $i=1,2,3,\dots$ gives an order to the clusters. The $c_{i}$ are chosen such that $c_1$ denotes the root cluster and $i<j$ if the first node of $c_i$ was attached to the memory tree before the first node of $c_j$. For technical reasons we might set $c_i=\emptyset$ for all $i$ that are larger than the number of clusters. We further denote the spin that is carried by all nodes of cluster $c_i$ by $m_i$, that is, $m_i=\pm 1$ or zero if $c_i = \emptyset$.
We denote the total magnetization, the sum of all spins, of cluster $c_i$ by $\mathcal{M}_{i}$. Since all spins of one cluster are equal, we have $\mathcal{M}_{i}=m_{i}|c_{i}|$, where $|c_{i}|$ denotes the number of nodes that belong to cluster $c_{i}$.
For the expectation value of the magnetization of the memory forest we find 
\begin{align}
	\langle x_{t=N}\rangle &= \langle M_{N}\rangle = \sum_{i=1}^{\infty} \langle \mathcal{M}_{i} \rangle = \sum_{i=1}^{\infty} \langle|c_i | \rangle \langle m_i\rangle \notag\\
	&= (2q-1)\langle |c_1|\rangle,
	\label{eq:rootclustersizeexpectation}
\end{align}
where $\langle \cdot \rangle$ denotes the expectation value over both, the construction of the random forest and the assignment of spins.
The only contribution comes from the root cluster, since the expectation value of the magnetization of all other clusters is zero, since they carry the spin $\pm 1$ with equal probability $1/2$.
Hence we connected the expectation value of the root cluster size with the first moment of the ERW that is known, cf. Eq. \eqref{eq:firstmoment}.
This leads to an exact expression for the expectation value of the root cluster size
\begin{align}
	\langle |c_1|\rangle= \frac{\Gamma(N+p)}{\Gamma(p+1)\Gamma(N)}.
	\label{eq:rootclustersizeexpectation2}
\end{align}
We can squeeze out even a little more. We call a cluster to be of the $k$th generation, if each path from the root to an arbitrary node of this cluster is interrupted by exactly $k$ deleted edges.
We are interested in the number of nodes in clusters of the first generation. Therefore we modify the coupling between ERWv1 and the percolation on RRTs, that is, we modify the construction of the memory tree.
We replace the first coin by a coin that can either show heads, tails or epsilon. We keep the probability $1-p$ to show tails, the probability to show heads is changed to $p-\varepsilon$ for some small parameter $\varepsilon>0$, and the first coin shows epsilon with probability $\varepsilon$.
The elephant is supposed to remember what it did in the past, when the first coin shows either heads or epsilon, such that the movement of the elephant will not be affected by these changes.
However we will change the construction of the memory tree. We delete an edge if the result of the corresponding first coin toss is either tails or epsilon. 
If the result is heads the edge is kept. We change nothing in the assignment of the spins, that is, the spin of node $i$ will still be $\sigma_{i}$, such that the magnetization $M_N$ still corresponds to the position of the elephant $x_{t=N}$.
All spins of nodes that belong to one cluster are still equal, but the assignments of spins to the clusters is not independent any more. 
More precisely the spin assigned to a cluster of the first generation will be coupled to the spin of the root cluster with probability $\varepsilon/(1-p+\varepsilon)$ and with probability $1-\varepsilon/(1-p+\varepsilon)$ it will be chosen independently.
Hence the expectation value of the magnetization is
\begin{align}
	&\langle M_{N}\rangle = (2q-1)\Big(\langle |c_1|\rangle+ \frac{\varepsilon}{1-p+\varepsilon} \sum_{i, c_{i} \text{ is $1$st generation}  } \langle |c_i|\rangle \notag \\
	&+ \left( \frac{\varepsilon}{1-p+\varepsilon} \right)^{2} \sum_{j, c_{j} \text{ is $2$nd generation }} \langle |c_{j}|\rangle + \mathcal{O}(\varepsilon^{3})\Big).
	\label{eq:magnetizationepsilon}
\end{align}
We have deleted each edge with probability $1-p+\varepsilon$, and according to Eq.~\eqref{eq:rootclustersizeexpectation2} we have
\begin{align}
	\langle |c_1|\rangle = \frac{\Gamma(N+p-\varepsilon)}{\Gamma(p-\varepsilon+1)\Gamma(N)}.
	\label{eq:rootclustersizeepsilon}
\end{align}
Inserting Eqs.~\eqref{eq:firstmoment} and \eqref{eq:rootclustersizeepsilon} into Eq.~\eqref{eq:magnetizationepsilon} and considering the first order in $\varepsilon$, that is, subtracting the left-hand side, dividing by $\varepsilon$ and taking the limit $\varepsilon \rightarrow 0$ we obtain
\begin{align}
	&\sum_{i, c_{i} \text{ is $1$st generation}} \langle |c_i|\rangle
	\label{eq:firstgenerationchild}\\
&= \frac{1-p}{p}\frac{\Gamma(N+p)}{\Gamma(p)\Gamma(N)}\Big( \Psi_{0}(N+p) - \Psi_{0}(p+1) \Big),\notag
\end{align}
where $\Psi_{0}$ denotes the digamma function.
Evaluating the terms of order $\varepsilon^{2}$ in Eq.~\eqref{eq:magnetizationepsilon} we can easily obtain an expression for the expectation value of the number of nodes in the second generation child clusters. Analogously one obtains the expectation value of the number of nodes in the $k$th generation child clusters for arbitrary $k\ge 1$.

So far we have used the first moment of the ERW to infer properties of clusters of percolation processes on RRTs.
In addition we can use the second moment of the ERW to observe further properties.
The second moment corresponds to the expectation value of the square of the magnetization of the memory forest.
\begin{align}
	\langle x^{2}_{t=N}\rangle &= \langle M_{N}^{2}\rangle = \langle \sum_{i=1}^{\infty} \mathcal{M}_{i}^{2}\rangle + 2 \sum_{j< k}\langle \mathcal{M}_{j} \mathcal{M}_{k}\rangle
	\label{eq:magnetizationsquared}\\ 
	&= \sum_{i=1}^{\infty}\langle m_{i}^{2} |c_{i}|^{2}\rangle + 2 \sum_{j < k }\langle m_{j} m_{k}\rangle \langle |c_{j}||c_{k}|\rangle \notag \\
	&= \sum_{i=1}^{\infty}\langle |c_{i}|^{2}\rangle = \frac{N}{2p-1} \left( \frac{\Gamma(N+2p)}{\Gamma(N+1)\Gamma(2p)}-1 \right),\notag
\end{align}
where we used that the spin assignment to the clusters is independent of the cluster structure and of other spins, hence $\langle m_{i}\rangle=0$ for all $i>1$. Therefore all mixed terms $\langle \mathcal{M}_{j}\mathcal{M}_{j}\rangle$ cancel. Furthermore $m_{i}^{2}=1$ when $m_{i}=\pm 1$. In the last line we inserted the result for the second moment from Eq.~\eqref{eq:secondmoment} with $\alpha=p$, cf. Eqs.~\eqref{eq:abbreviations} and \eqref{eq:parameterrelation}. Thus we found the expectation value of the sum of the cluster sizes squared.

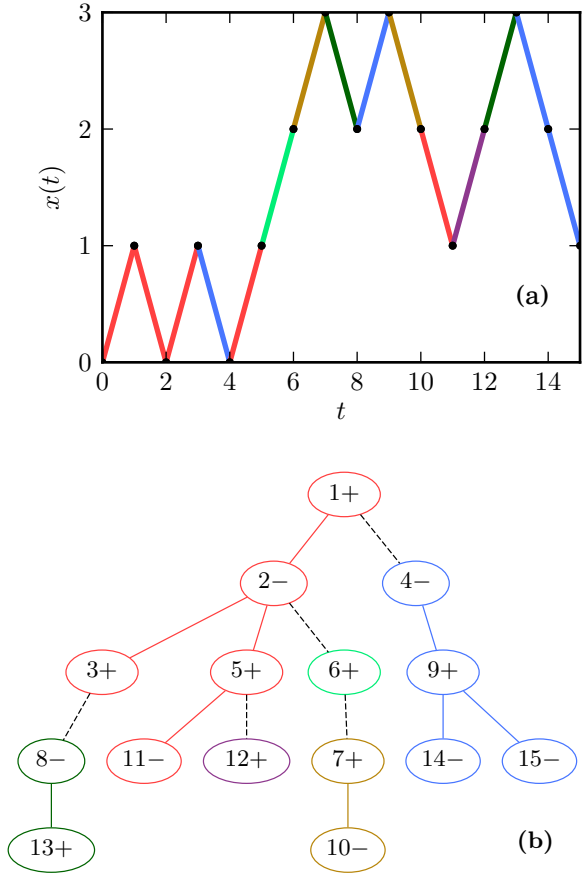
\begin{figure}
	\includegraphics{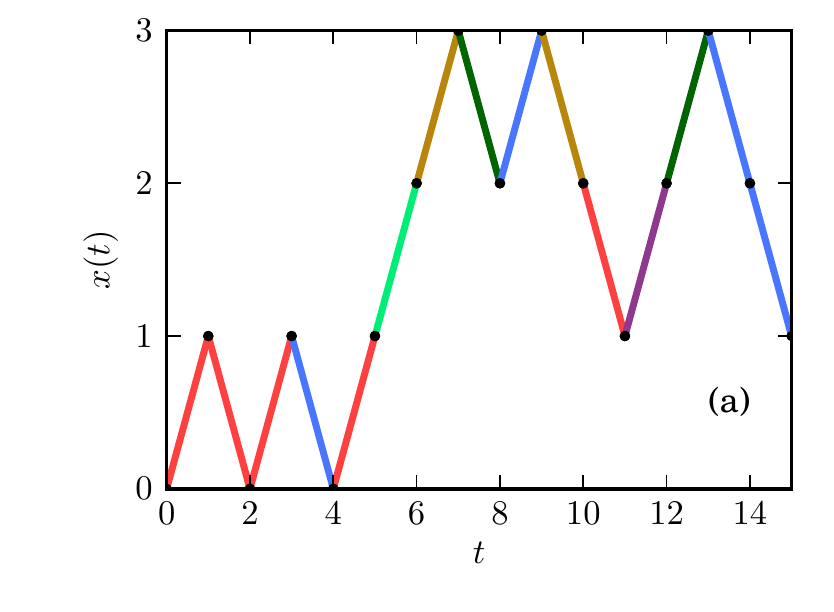}
\def\svgwidth{0.45\textwidth}
\begingroup%
  \makeatletter%
  \providecommand\color[2][]{%
    \renewcommand\color[2][]{}%
  }%
  \providecommand\transparent[1]{%
    \renewcommand\transparent[1]{}%
  }%
  \providecommand\rotatebox[2]{#2}%
  \ifx\svgwidth\undefined%
    \setlength{\unitlength}{468bp}%
    \ifx\svgscale\undefined%
      \relax%
    \else%
      \setlength{\unitlength}{\unitlength * \real{\svgscale}}%
    \fi%
  \else%
    \setlength{\unitlength}{\svgwidth}%
  \fi%
  \global\let\svgwidth\undefined%
  \global\let\svgscale\undefined%
  \makeatother%
  \begin{picture}(1,0.70940171)%
    \put(0,0){\includegraphics[width=\unitlength]{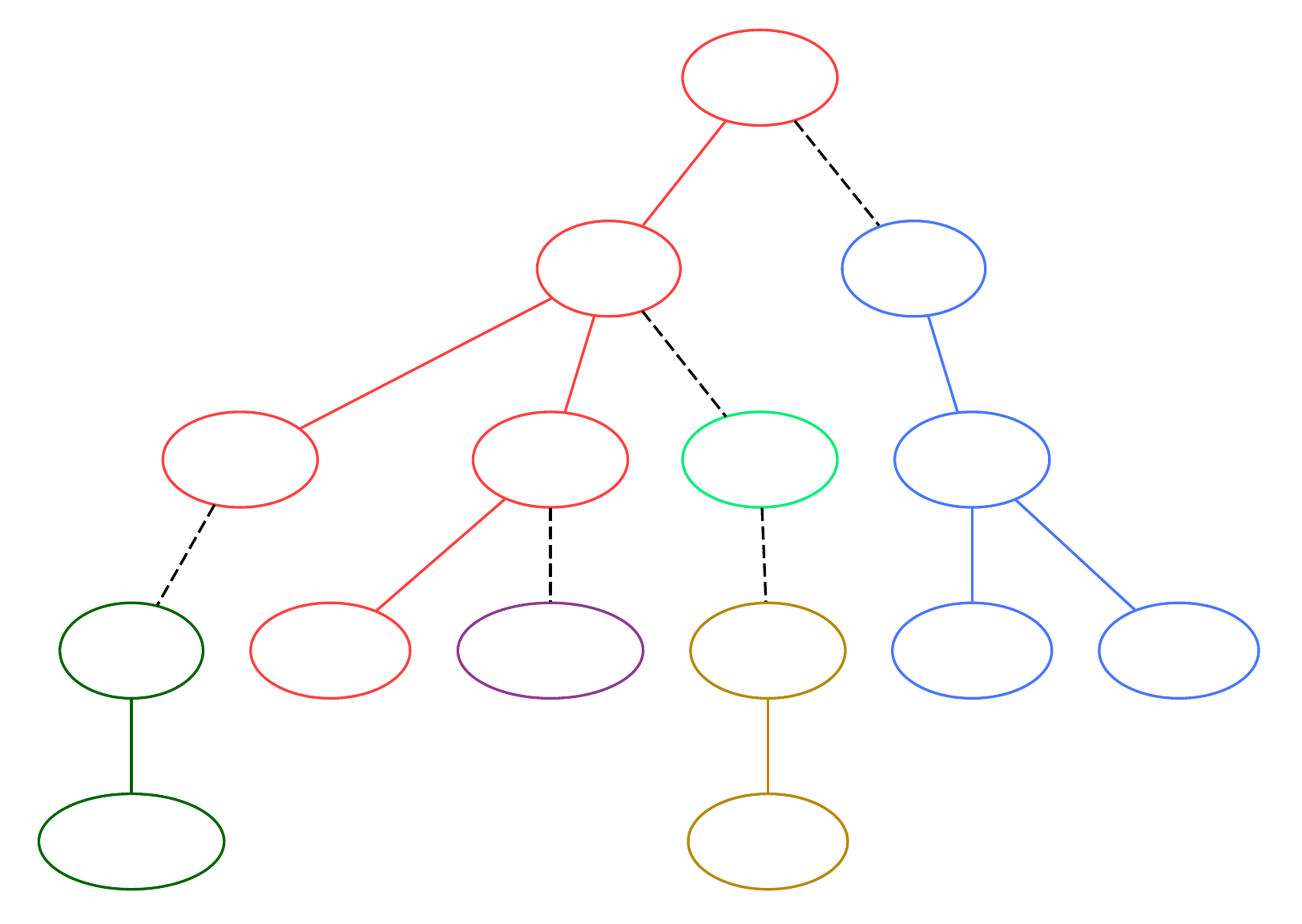}}%
    \put(0.5836935,0.64140503){\makebox(0,0)[b]{\smash{$1 +$}}}%
    \put(0.46759966,0.49476019){\makebox(0,0)[b]{\smash{$2 -$}}}%
    \put(0.70182407,0.49476019){\makebox(0,0)[b]{\smash{$4 -$}}}%
    \put(0.18449364,0.34811534){\makebox(0,0)[b]{\smash{$3 +$}}}%
    \put(0.42279152,0.34811534){\makebox(0,0)[b]{\smash{$5 +$}}}%
    \put(0.5836935,0.34811534){\makebox(0,0)[b]{\smash{$6 +$}}}%
    \put(0.10098755,0.20147051){\makebox(0,0)[b]{\smash{$8 -$}}}%
    \put(0.74663221,0.34811534){\makebox(0,0)[b]{\smash{$9 +$}}}%
    \put(0.2537426,0.20147051){\makebox(0,0)[b]{\smash{$11 -$}}}%
    \put(0.42279152,0.20147051){\makebox(0,0)[b]{\smash{$12 +$}}}%
    \put(0.5898037,0.20147051){\makebox(0,0)[b]{\smash{$7 +$}}}%
    \put(0.5898037,0.05482567){\makebox(0,0)[b]{\smash{$10 -$}}}%
    \put(0.10098755,0.05482567){\makebox(0,0)[b]{\smash{$13 +$}}}%
    \put(0.74663221,0.20147051){\makebox(0,0)[b]{\smash{$14 -$}}}%
    \put(0.90549746,0.20147051){\makebox(0,0)[b]{\smash{$15 -$}}}%
    \put(0.868,0.05641978){\color[rgb]{0,0,0}\makebox(0,0)[lb]{\textbf{(b)}}}%
  \end{picture}%
\endgroup%
	\caption{(color online) Trajectory (a) and memory forest (b) of the same realization of ERWv2. Deleted edges are indicated by dashed lines. Time steps, or nodes, that belong to one memory component, that is one cluster, are drawn with the same color. $q=1/2$, $p=0.7$.\label{fig:erwv2}}
\end{figure}

We successfully used ERWv1 to study percolation on RRTs. We can also use the ERWv2.
The construction of the memory tree and forest remains almost the same apart from differences in the spin assignments.
Spins of one cluster are not identical but spins of neighboring nodes have opposite spins.
This is because the elephant does always the opposite when it decides to remember what it did in the past.
The spin of the root of each cluster is determined by the toss of the second coin. 
All other spins of the same cluster follow from the tree structure. In Fig.~\ref{fig:erwv2} we show one realization of the ERWv2 until time $t=15$ and the corresponding memory forest.

Given that the root of a cluster $c_{i}$ has spin $+1$ and the size of the cluster is at least two, we have $\langle \mathcal{M}_{i}=0\rangle $ which we prove in Appendix \ref{app:a}.
The same result holds when the root of the cluster has spin $-1$.
For all clusters except the root cluster, $\langle \mathcal{M}_{i} =0\rangle$ remains true also if the cluster has size one, since then it carries spin $\pm 1$ with equal probability $1/2$.
Hence
\begin{align}
	\langle x_{t=N} \rangle &= \langle M_{N}\rangle = \sum_{i=1}^{\infty} \langle \mathcal{M}_{i}\rangle = \langle \mathcal{M}_{1}\rangle
	\label{eq:magnetizationsecondversion}\\
	&= Pr(|c_{1}|=1)\langle m_{1}\rangle = Pr(|c_{1}|=1) (2q-1).\notag
\end{align}
Thus we obtain with Eq.~\eqref{eq:firstmoment} the probability that the root is isolated
\begin{align}
	Pr(|c_{1}|=1) = \frac{\Gamma(N-p)}{\Gamma(N)\Gamma(1-p)}.
	\label{eq:probisolatedroot}
\end{align}

To infer information on the percolation process on RRTs we can consider a modified version of the ERW.
As in the ERW the elephant starts at $x_{0}=0$ and goes right in the first step with probability $q$ and left with probability $1-q$.
In each following step it chooses a time $t'$ from the past with uniform probability. Then
\begin{align}
	\sigma_{t+1}= \begin{cases}
		\phantom{-+1} \sigma_{t'} &\text{with probability } \tilde{p}\\
		\phantom{+1} -\sigma_{t'} &\text{with probability } 1-\tilde{p} -r\\
		\phantom{-\sigma_{t'}} +1 &\text{with probability } r.
	\end{cases}
	\label{eq:incrementmodifiederw}
\end{align}
That means it either does the same as it did in the past, or the opposite, or it goes right, independent on the past.
Clearly $\tilde{p}$, $1-\tilde{p}-r$ and $r$ must all be probabilities, hence $0\le r+\tilde{p}\le 1$ must be satisfied.
This model is only a slight modification of the original ERW that is recovered for $r=0$.
We will refer to this process as the skew elephant random walk (SERW).
We can apply the same method as \cite{ST04} to calculate the first moment.
The probability to go right in step $t+1$, given the past, is
\begin{align}
	Pr(\sigma_{t+1}&=+1|\sigma_{1}, \dots, \sigma_{t})= r + \tilde{p}Pr(\sigma_{t'}=1|\sigma_{1}, \dots, \sigma_{t}) \notag \\
	&+(1-\tilde{p}-r)Pr(\sigma_{t'}=-1|\sigma_{1}, \dots, \sigma_{t}).
	\label{eq:probabilityright}
\end{align}
Suppose the elephant moved $k$ times to the right and $t-k$ times to the left until time $t$. Then clearly $x_t=k-(t-k)=2k-t$ and consequently $k=(x_{t}+t)/2$. Hence
\begin{align}
	Pr(\sigma_{t'}=+1|\sigma_{1}, \dots, \sigma_{t})&= \frac{k}{t}= \frac{t+x_{t}}{2t}, \notag \\
	Pr(\sigma_{t'}=-1|\sigma_{1}, \dots, \sigma_{t})&= \frac{t-k}{t}= \frac{t-x_{t}}{2t}.
	\label{eq:probabilityrememberrightleft}
\end{align}
Inserting this expression into Eq.~\eqref{eq:probabilityright} we find
\begin{align}
	Pr(\sigma_{t+1}=+1|\sigma_{1}, \dots, \sigma_{t})= \frac{r+1}{2} + \frac{2\tilde{p}+r-1}{2}\frac{x_{t}}{t}.
	\label{eq:probabilityright2}
\end{align}
From this expression we can calculate the expectation value of the increment as
\begin{align}
	\langle \sigma_{t+1}|\sigma_{1}, \dots, \sigma_{t}\rangle =& +1\cdot Pr(\sigma_{t+1}=+1|\sigma_{1}, \dots, \sigma_{t}) \notag \\
	&-1\cdot (1-Pr(\sigma_{t+1}=+1|\sigma_{1}, \dots, \sigma_{t}))\notag \\
	=& r + (2\tilde{p}+r-1)\frac{x_{t}}{t}.
	\label{eq:conditionalexpectationincrement}
\end{align}
Averaging over all realizations until time $t$ we find
\begin{align}
	\langle \sigma_{t+1}\rangle = r + (2\tilde{p}+r-1)\frac{\langle x_{t} \rangle }{t}.
	\label{eq:expectationincrement2}
\end{align}
Taking the expectation value of Eq.~\eqref{eq:timeevolution} we find from this expression
\begin{align}
	\langle x_{t+1}\rangle &= r + \Big(1+\frac{\gamma}{t}\Big)\langle x_{t}\rangle, \text{ with}\\
	\gamma &= 2\tilde{p} +r -1.
	\label{eq:gamma}
\end{align}
With the initial value $\langle x_{1}\rangle=2q-1$ the explicit expression for $\langle x_{t}\rangle$ follows
\begin{align}
	\langle x_{t}\rangle = \Big(2q-1-\frac{r}{1-\gamma}\Big)\frac{\Gamma(t+\gamma)}{\Gamma(t)\Gamma(\gamma+1)} + \frac{rt}{1-\gamma}.
	\label{eq:firstmomentgeneralized}
\end{align}
For completeness we will also calculate the second moment. From Eq.~\eqref{eq:timeevolution} we find
\begin{align}
	x_{t+1}^{2}=x_{t}^{2} + 2\sigma_{t+1}x_{t} + \sigma_{t+1}^{2}.
	\label{eq:secondmomentskew1}
\end{align}
Taking into account that $\sigma_{t+1}^{2}=1$ and using Eq.~\eqref{eq:conditionalexpectationincrement} we find for the second moment
\begin{align}
	\langle x_{t+1}^{2}\rangle =& 1+ 2r \langle x_{t}\rangle + \Big(1+  \frac{2\gamma}{t} \Big) \langle x_{t}^{2}\rangle
	\notag
	\\
	=&1 + 2r\Big(2q-1-\frac{r}{1-\gamma}\Big)\frac{\Gamma(t+\gamma)}{\Gamma(t)\Gamma(\gamma+1)} 
	\notag
	\\
	&+ \frac{2r^{2}t}{1-\gamma}+ \Big(1+  \frac{2\gamma}{t} \Big) \langle x_{t}^{2}\rangle,
	\label{eq:secondmomentskew2}
\end{align}
where we inserted Eq.~\eqref{eq:firstmomentgeneralized}. With the initial condition $x_{1}^{2}=1$ we find the explicit formula
\begin{align}
	&\langle x_{t}^{2}\rangle =\frac{\Gamma(\gamma+t)}{\Gamma(\gamma+1)\Gamma(t)}\big(-2r(t+1) \big) \bigg(\frac{2q-1}{\gamma-1} + \frac{r}{(\gamma-1)^{2}} \bigg)
	\notag
	\\
	&+\frac{2\Gamma(2\gamma+t)}{\Gamma(2\gamma+1)\Gamma(t)}\bigg(\frac{2r(2q-1)}{\gamma-1} + \frac{r^{2}(3\gamma-2)}{(\gamma-1)^{2}(2\gamma-1)}
	\label{eq:secondmomentskew3}
	\\
	&+ \frac{\gamma}{2\gamma-1} \bigg)
	+\frac{r^{2}t^{2}}{(\gamma-1)^{2}} - \frac{t}{2\gamma-1}+\frac{r^{2}t}{(\gamma-1)^{2}(2\gamma-1)}.
	\notag
\end{align}

Similar to the ERW we can reformulate also the SERW. For $\tilde{p}=1-\tilde{p}-r$ the SERW is just an ordinary random walk where the elephant goes right with probability $\tilde{p}+r$ and left with probability $1-\tilde{p}-r$ in all time steps starting from $t=1$ and independent on the past. For completeness we will also give SERWv1 although we will not use it further.
This is the case when $\tilde{p}>1-\tilde{p}-r$. In SERWv1 the first step is the same as before and for any later time
\begin{align}
	\sigma_{t+1}= \begin{cases}\sigma_{t'} &\text{with probability }p \\
		+1  &\text{with probability } 1-\tilde{p} \\
		-1 &\text{with probability } 1-\tilde{p}-r. 
	\end{cases}
	\label{eq:incrementserwv1}
\end{align}
This formulation is equivalent to SERW when
\begin{align}
	p=2\tilde{p}+r-1=\gamma.
	\label{eq:equivalenceserwv1}
\end{align}
The last case, we call it SERWv2, appears for $\tilde{p}<1-\tilde{p}-r$ and differs from the previous only in the increment which now reads
\begin{align}
	\sigma_{t+1}= \begin{cases}-\sigma_{t'} &\text{with probability }p \\
		+1 &\text{with probability } \tilde{p}+r \\
		-1 &\text{with probability } \tilde{p}. 
	\end{cases}
	\label{eq:incrementserwv2}
\end{align}
This formulation is equivalent to SERW when
\begin{align}
	p=-2\tilde{p}-r+1=-\gamma.
	\label{eq:equivalenceserwv2}
\end{align}
Considering the SERWv2 we can infer information on the number of clusters of size one from its respective memory tree. For each time step in which the elephant is remembering and doing the opposite as in the past, the corresponding edge is kept and in case the elephant goes right or left, the corresponding edge is removed. Hence vertexes that are connected by an edge carry opposite spin and the spins of vertexes that do not belong to the same cluster are independent of each other.
As we have argued before, all clusters are random recursive trees themselves and if their size is larger than one, the expected magnetization is zero independent on the spin of the root, cf. Appendix \ref{app:a}. The expectation value of the spin of the root of each child cluster is $r/(2\tilde{p}+r)=r/(1-p)$.
Hence we find the expectation of the magnetization
\begin{align}
	\langle x_{t=N}\rangle=&\langle M_{N}\rangle = \sum_{i=1}^{\infty} \langle \mathcal{M}_{i}\rangle = (2q-1)Pr(|c_1|=1)
	\notag\\
	&+\sum_{i=2}^{\infty}\frac{r}{1-p}Pr(|c_{i}|=1)
	\notag\\
	=&(2q-1)\frac{\Gamma(N-p)}{\Gamma(N)\Gamma(1-p)} 
	\notag \\
	&+ \frac{r}{1-p}\langle \# i>1: |c_{i}|=1\rangle
	\notag \\
	=& (2q-1 - \frac{r}{p+1})\frac{\Gamma(N-p)}{\Gamma(N)\Gamma(1-p)}+\frac{rN}{1+p}.
	\label{eq:clustersofsizeone}
\end{align}
where we have plugged Eq.~\eqref{eq:probisolatedroot} into the first line and the last line comes from Eq.~\eqref{eq:firstmomentgeneralized} with $\gamma=-p$, cf. Eq.~\eqref{eq:equivalenceserwv2}. From this expression we obtain the expectation value of the number of clusters different from the root cluster of size one
\begin{align}
	\langle \# i>1: |c_{i}|=1\rangle= \frac{1-p}{1+p}\left[ \frac{-\Gamma(N-p)}{\Gamma(N)\Gamma(1-p)} + N \right].
	\label{eq:clustersofsizeone2}
\end{align}
Adding the probability that the root cluster is of size one, cf. Eq.~\eqref{eq:probisolatedroot}, we find the expectation value of the number of clusters that are of size one
\begin{align}
	\langle \# i\ge1 : |c_{i}|=1\rangle = \frac{2p}{1+p} \frac{\Gamma(N-p)}{\Gamma(N)\Gamma(1-p)} + N \frac{1-p}{1+p}.
	\label{eq:clustersofsizeone3}
\end{align}

In the following we use a process introduced in \cite{KHL10} that we call the lazy elephant random walk (LERW) due to the fact that in this model the elephant is allowed to sleep, that means it is allowed to stay at the same position for one time step. In the first step the elephant still moves right with probability $q$ and left with probability $1-q$.
The position after all proceeding steps is determined by Eq.~\eqref{eq:timeevolution} with
\begin{align}
	\sigma_{t+1} = \begin{cases}
		\phantom{-}\sigma_{t'} &\text{with probability } p\\
		-\sigma_{t'} &\text{with probability } 1-p-s\\
		\phantom{-}0 &\text{with probability } s,
	\end{cases}
	\label{eq:updatelazyelephant}
\end{align}
where $t'$ is randomly chosen from $\left\{ 1, \dots, t \right\}$ and the system parameters $s, p$ must be chosen in such a way that all appearing probabilities are between zero and one.

The first and second moments that have been calculated in \cite{KHL10} are

\begin{align}
	\langle x_{t}\rangle &= (2q-1)\frac{\Gamma(t+2p+s-1)}{\Gamma(t)\Gamma(2p+s)},
	\label{eq:explicitfirstmomentlazy}
	\\
	\langle x_{t}^{2}\rangle &= \frac{1}{4p+3s-3}\left( \frac{\Gamma(4p+2s+t-2)}{\Gamma(4p+2s-2)\Gamma(t)} - \frac{\Gamma(1-s+t)}{\Gamma(1-s)\Gamma(t)} \right).
	\label{eq:secondmomentresultlazy}
\end{align}
For large $t$ the leading behavior is
\begin{align}
	\langle x_{t}^{2}\rangle \approx \begin{cases}
		\frac{t^{1-s}}{(3-4p-3s)\Gamma(1-s)} &\text{for }4p+3s-3<0\\
		\frac{1}{\Gamma(1-s)}t^{t-s}\ln(t) &\text{for } 4p+3s-3=0\\
		\frac{t^{4p+2s-2}}{(4p+3s-3)\Gamma(4p+2s-2)} &\text{for } 4p+3s-3>0.
	\end{cases}
	\label{eq:secondmomentassymplazy}
\end{align}
Depending on the choice of parameters we can obtain all exponents between zero and two. Thus the model can show subdiffusion as well as superdiffusion. Note that Eq.~\eqref{eq:meansquareddisplacement} is recovered for $s=0$.

To obtain the second moment of the root cluster size we set the parameter $s=1-p$. That means the elephant either repeats what it did in the past or it sleeps. We can construct again the memory forest of the elephant by keeping the edges whenever the elephant decides to do the same as in the past and by deleting the corresponding edge whenever the elephant decides to sleep. By this construction the spins of the root cluster are all either plus or minus one and the spins of all other clusters are zero. Hence we find for the square of the magnetization
\begin{align}
	\langle x^{2}_{t=N}\rangle = \langle M_{N}^{2}\rangle = \langle |c_1|^{2}\rangle,
	\label{eq:magnetizationsquaredlazy}
\end{align}
and with Eq.~\eqref{eq:secondmomentresultlazy} we find an exact expression for the second moment of the root cluster size
\begin{align}
	\langle |c_{1}|^{2}\rangle= \frac{1}{p}\left( \frac{\Gamma(2p+N)}{\Gamma(2p)\Gamma(N)} - \frac{\Gamma(p+N)}{\Gamma(p)\Gamma(N)} \right).
	\label{eq:rootclustersecondmoment}
\end{align}
Using certain $\varepsilon$-modifications of the coupling between LERW and percolation on RRTs similar to the one used before one can obtain correlations of the root cluster size with the number of nodes in first generation child clusters
\begin{align}
	&\langle |c_1|(\# \text{nodes in $1$st generstion child clusters})\rangle
	\label{eq:correlationrootfirstgen}\\
	&=\frac{1-p}{p} \bigg[\frac{\Gamma(2p+N)}{\Gamma(2p)\Gamma(N)}\Big( -\frac{1}{p} - \Psi_{0}(2p)+  \Psi_{0}(2p+N) \Big)
	\notag \\
	&\phantom{=} + \frac{1}{p} \frac{\Gamma(p+N)}{\Gamma(p)\Gamma(N)} \bigg],
	\notag
\end{align}
as well as the expectation value of the number of nodes in first generation child clusters squared
\begin{align}
	&\langle (\# \text{nodes in $1$st generation child clusters})^{2}\rangle
	\label{eq:firstgensecondmoment} \\
	&=\frac{(1-p)^{2}}{p^{2}} \bigg[ \frac{\Gamma(2p+N)}{\Gamma(2p)\Gamma(N)}\Big(\frac{8}{p} +\frac{1}{1-p} + 6 \Psi_{0}(2p) 
	\notag
	\\
	&\phantom{=} -6\Psi_{0}(2p+N) +2p \big(\Psi_{0}^{2}(2p)+ \Psi_{0}^{2}(2p+N) 
	\notag
	\\
	&\phantom{=} -2 \Psi_{0}(2p)\Psi_{0}(2p+N) - \Psi_{1}(2p) + \Psi_{1}(2p+N)   \big) \Big)
	\notag
	\\
	&\phantom{=} +\frac{\Gamma(p+N)}{\Gamma(p)\Gamma(N)}\Big( -\frac{8}{p} - \frac{1}{1-p} + \frac{p}{1-p}\big(\Psi_{0}(p)
	\notag
	\\
	&\phantom{=}-\Psi_{0}(p+N) \big) +2\Psi_{0}(p) -2\Psi_{0}(p+N)
	\notag
	\\
	&\phantom{=} -2p\Psi_{0}(p)\Psi_{0}(p+N)\Big)\bigg],
	\notag
\end{align}
where $\Psi_{k}$ denotes the polygamma function.
Details of the calculation are given in Appendix \ref{app:b}.

\section{Comparison with previous results\label{sec:6}}

The main quantity that has been discussed in \cite{Bertoin14,Baur14, BB14} is the size of the root cluster in certain limits, where $N\rightarrow \infty$ and $p\rightarrow 1$. In this regime the root cluster is the largest cluster with high probability. Also the sizes of the next largest clusters have been investigated in these works.
The result may depend on the precise way the aforementioned limits are assumed. Therefore we need to state the limit more precisely and there are several ways to do it. One limit that was considered in \cite{Bertoin14} is
\begin{align}
	&N \rightarrow \infty,
	\notag\\
	&p(N)= 1 -\lambda/\ln(N).
	\label{eq:supercritical}
\end{align}
This limit is called supercritical in the literature. The first part of the main result of \cite{Bertoin14} is that in this limit
\begin{align}
	\frac{1}{N}|c_{1}| \rightarrow \exp(-\lambda),
	\label{eq:limitbertoin1}
\end{align}
where the convergence is in probability.
We obtained an exact expression for the expectation value of the root cluster size for finite $N$, cf. Eq.~\eqref{eq:rootclustersizeexpectation2}. Evaluating the root cluster size divided by $N$ in the limit \eqref{eq:supercritical} we find
\begin{align}
	\frac{1}{N}\langle |c_{1}|\rangle \rightarrow \exp(-\lambda).
	\label{eq:meanvalue}
\end{align}
We can also calculate the variance
\begin{align}
	&\lim_{N\rightarrow \infty}\langle (\frac{1}{N}|c_{1}|-\exp(-\lambda))^{2}\rangle
	\notag \\
	&= \lim_{N\rightarrow \infty} \frac{1}{N^{2}} \langle |c_{1}|^{2}\rangle - \exp(-2\lambda).
	\label{eq:variancerootclustersize1}
\end{align}
As we have also the exact expression of the second moment of the root cluster size, cf. Eq.~\eqref{eq:rootclustersecondmoment}, we can calculate the limit
\begin{align}
	\lim_{N\rightarrow \infty} \frac{1}{N^{2}} \langle |c_1|^{2}\rangle = \exp(-2\lambda). 
	\label{eq:limitrootclustersizesquared}
\end{align}
Hence the limit \eqref{eq:limitbertoin1} is also achieved in mean square convergence, which implies convergence in probability.

The second part of the main result of \cite{Bertoin14} deals with the sizes of the next largest clusters after the root cluster. It is stated that if $C_{1}, \dots, C_{l}$ denote the $l$ next largest clusters then
\begin{align}
	\left(\frac{\ln N}{N}C_{1}, \dots, \frac{\ln N}{N} C_{l}  \right) \rightarrow (x_{1}, \dots, x_{l}),
	\label{eq:result2bertoin}
\end{align}
where the convergence is in distribution in the limit \eqref{eq:supercritical} and $x_{1}>x_{2}>\dots$ denote the sequence of atoms of a Poisson random measure on $(0, \infty)$ with intensity $\lambda\exp(-\lambda)x^{-2}\diff x$.
This result clearly outruns our results as it gives detailed information over the next largest clusters. However as it was argued in \cite{Baur14} the next largest clusters are of first generation with high probability. Hence our result Eq.~\eqref{eq:firstgenerationchild} on the number of nodes in first generation child clusters is related to it.
Considering the limit \eqref{eq:supercritical} we find
\begin{align}
	\frac{\ln N}{N} \langle \# \text{nodes in $1$st gen. child clusters}\rangle
	\notag \\
	\rightarrow \lambda \exp(-\lambda).
	\label{eq:limitfirstgen}
\end{align}
Furthermore we find from Eq.~\eqref{eq:firstgensecondmoment} in the limit \eqref{eq:supercritical}
\begin{align}
	\bigg(\frac{\ln N}{N}\bigg)^{2} \langle (\# \text{nodes in $1$st gen. child clusters})^{2}\rangle
	\notag \\
	\rightarrow 2\lambda^{2} \exp(-2\lambda),
	\label{eq:limitfirstgensquared}
\end{align}
such that the variance of the scaled quantity $\frac{\ln N}{N} (\# \text{nodes in $1$st gen. child clusters})$ converges to $\lambda^{2}\exp(-2\lambda)$. That means the scaled number of nodes in first generation child clusters is still fluctuating in the limit \eqref{eq:supercritical}, which is in accordance with \cite{Bertoin14}. 

The main results of \cite{Baur14} generalize the findings of \cite{Bertoin14} as \cite{Baur14} deals with more general limits of the form
\begin{align}
	N\rightarrow \infty, \qquad
	p(N) \rightarrow 1,
	\label{eq:limitbaur}
\end{align} 
where \eqref{eq:supercritical} is called supercritical, the case $1/\ln N \ll 1- p(N) \ll 1$ is called weakly supercritical and the case $0<1-p(N)\ll 1/\ln N$ is called strongly supercritical. The results of \cite{Baur14} deal with the size of clusters of all generations. For the root cluster size the result is
\begin{align}
	\frac{1}{N^{p(N)}}|c_{1}|\rightarrow 1,
	\label{eq:limitrootclusterbaur}
\end{align}
where the limit is in distribution. Also in these limits our results Eqs.~\eqref{eq:rootclustersizeexpectation2} and \eqref{eq:rootclustersecondmoment} lead to the limit \eqref{eq:limitrootclusterbaur} in mean square convergence.

Some authors investigated the cutting of random recursive trees until the root is isolated \cite{MM74, IM07, KP14}. In this process a randomly chosen edge is removed and afterwards only the connected component that contains the root is considered. The removing of edges is repeated until the root is isolated. Our results are not so much connected to this process, however we give the probability that the root is isolated in percolation on RRTs, cf. Eq.~\eqref{eq:probisolatedroot}, which is in some sense a complementary consideration.

\section{A percolation view on anomalous diffusion\label{sec:7}}

We are using the analogy between ERW and percolation on RRTs to come to a deeper understanding of superdiffusion in ERW.
For this purpose the study of the root cluster size is essential. 
According to Eq.~\eqref{eq:rootclustersizeexpectation2} the expectation value of the root cluster size behaves for fixed $p$ and large $N$ as
\begin{align}
	\langle |c_{1}|\rangle\approx \frac{1}{\Gamma(p+1)} N^{p}.
	\label{eq:rootclusterexpectationlargeN}
\end{align}
Considering also the second moment \eqref{eq:rootclustersecondmoment} one easily checks that
\begin{align}
	\frac{|c_{1}|}{N^p} \rightarrow \frac{1}{\Gamma(p+1)}
	\label{eq:limitrootclustersize}
\end{align}
in mean square convergence. Hence for large $N$, the leading behavior of the root cluster size is
\begin{align}
	|c_{1}|\approx \frac{1}{\Gamma(p+1)} N^{p}.
	\label{eq:leadingbehaviorrootclustersize}
\end{align}
 From Eq.~\eqref{eq:rootclustersecondmoment} we find also the leading behavior of the second moment of the root cluster size for large $N$.
\begin{align}
	\langle |c_{1}|^{2} \rangle \approx \frac{1}{p \Gamma(2p)}N^{2p}.
	\label{eq:rootclustersecondmomentlargeN}
\end{align}
We are considering now ERW in its formulations ERWv1 and ERWv2. In ERWv2 there is no anomalous diffusion.

In the memory forest of ERWv1 the spins of all nodes belonging to one cluster are equal. According to Eq.~\eqref{eq:magnetizationsquared} we have
\begin{align}
	\langle x_{t}^{2}\rangle = \langle |c_1|^{2}\rangle + R,
	\label{eq:secondmomentremainder}
\end{align}
where the remainder is
\begin{align}
	R= \sum_{i=2}^{\infty} \langle |c_{i}|^{2}\rangle.
	\label{eq:remainder}
\end{align}
Taking into account the asymptotic second moment of the root cluster size \eqref{eq:rootclustersecondmomentlargeN} with $t=N$ we find
\begin{align}
	\langle x_{t}^{2}\rangle \approx \frac{1}{p\Gamma(2p)} t^{2p} + R.
	\label{eq:variancelargeN}
\end{align}
Hence for $p>1/2$ the superdiffusive behavior of ERWv1 can be explained solely by the root cluster size of the memory forest of ERWv1. That means the leading contribution to the second moment comes exclusively from the memory component that contains the first step.

For $p < 1/2$ also the other cluster sizes are important. The root cluster size squared grows sub-linearly with the system size with large probability, cf. Eq.~\eqref{eq:leadingbehaviorrootclustersize}. Each other clusters is likely to be smaller than the root cluster. Thus the leading contribution in \eqref{eq:magnetizationsquared} does not come from a single cluster, but from all clusters. Since the number of clusters is growing linearly in $t$ with high probability, the second moment also grows linearly in time and hence normal diffusion is observed.

We also want to come to a deeper understanding of subdiffusion in the LERW model \cite{KHL10}.
Therefore we are going to modify the construction of the memory forest once more.
Each time the elephant is remembering a time in the past it was not sleeping and it decides to sleep in this step, the corresponding edge is deleted. That means these edges are deleted with probability $s$.
If the elephant is remembering a step in the past when it was sleeping, it will sleep with probability one.
However we will delete the corresponding edge also in this case with probability $s$. Thus each edge is kept with probability $1-s$ and deleted with probability $s$.

We observe that the spins of all nodes that do not belong to the root cluster are zero. These nodes correspond to the steps the elephant is sleeping.
On the other hand, the spins of all nodes of the root cluster are non-zero and they are distributed as the steps of ERW with parameter
\begin{align}
	\tilde{p}=\frac{p}{1-s}.
	\label{eq:parameterERW}
\end{align}
Since the root cluster is an RRT of random size itself we can construct a realization of LERW at time $t$ as follows.
Choose a random time $T$, we call it the effective time, from a distribution equal to the distribution of the root clusters size for percolation on a RRT of size $t$ and with percolation parameter $1-s$.
Then perform an ERW with parameter $\tilde{p}=p/(1-s)$ until time $T$.
That means that LERW can be seen as an ERW with an effective time $T$, that is a random variable. For $t\rightarrow \infty$ the leading behavior of $T$ is deterministic, according to Eq.~\eqref{eq:leadingbehaviorrootclustersize}
\begin{align}
	T \approx \frac{1}{\Gamma(2-s)}t^{1-s}.
	\label{eq:effectivetimeasymp}
\end{align}
Replacing the time in Eq.~\eqref{eq:meansquareddisplacement} by the effective time \eqref{eq:effectivetimeasymp} and taking Eq.~\eqref{eq:parameterERW} into account we find the leading behavior of LERW \eqref{eq:secondmomentassymplazy}.

One can imagine to choose the root cluster not according to ERW but following some other process. If we choose a ballistic motion, that means all spins of the root cluster are plus one, with the effective time as root cluster size we appear at the process investigated in \cite{HKL14}. A generalization of LERW is obtained when the root cluster is chosen according to SERW.

\section{Conclusions\label{sec:8}}

In this paper we revealed a surprising connection between percolation on random recursive trees and the elephant random walk or similar models.
We used a stochastic coupling between these models to obtain exact results for the expectation values of certain properties of the percolation process on RRTs.
These results allowed us to reobtain certain limiting results on the root cluster size for percolation on RRTs known in the literature. The results of this paper are stronger than previous ones as we have proved mean square convergence, which implies the previously proved convergence in probability.

All results on percolation on RRTs in this paper rely in principle on recursion relations of expectation values of certain quantities. We want to remark that in principle we do not need to define any random walk to obtain these results. Not even the assignment of a spin to each node is necessary. However the calculations become much more intuitive using this formulation and it directly gives an idea how certain quantities can be calculated. Furthermore it is interesting on its own that percolation on RRTs is related to random walks that show anomalous diffusion.

In the opinion of the author, reformulations of the ERW model introduced in this paper, as well as the correspondence with RRTs, lead to a deeper understanding of ERW-like models.
In particular we gave an demonstrative explanation of subdiffusion present in LERW by introducing an effective time. We also found that superdiffusion in ERW is due to one giant memory component only.
We further introduced a new generalization of the ERW, the skew elephant random walk, for which we calculated the first and the second moment.

The author believes that the correspondence between ERW-like models and percolation on RRTs gives a powerful tool to solve even more problems that might appear in the study of percolation on RRTs. Furthermore, future work might profit from the equivalence of these models also in the other direction, where one might infer information on ERW-like models from what is known or might be found in the future about percolation on RRTs. 

\begin{appendix}
\section{Magnetization of random recursive trees with anti-aligned spins\label{app:a}}
We consider RRTs of size $N$ where the root node has spin $\sigma_{1}=+1$ and for neighboring nodes $i,j$ the spins have opposite signs $\sigma_{i}=-\sigma_{j}$. We prove that the expectation value of the magnetization of the RRT is one for $N=1$ and zero for $N\ge 2$.
	The case $N=1$ is trivial. The case $N=2$ is also clear since there is only one possible RRT of size two, for which we have $\sigma_{1}=+1, \sigma_{2}=-1$.
	Let us denote the set of all possible RRTs of size $N$ by $T_{N}$. Once we have chosen a realization $\tau \in T_{N}$, then the magnetization $M_{N}(\tau)$ is already determined.
	For $N>2$ we will define a bijective map $f:T_{N} \rightarrow T_{N}$ that satisfies
	\begin{align}
		M_{N}(\tau)=-M_{N}(f(\tau)).
		\label{eq:bijection}
	\end{align}
	The map $f$ interchanges the child nodes of the first two nodes. That means if in the RRT $\tau$ any node $k>2$ is a child of the root node, it will be a child of the second node in $f(\tau)$ and if any node $k>2$ is a child of the second node in $\tau$, it will be a child of the root node in $f(\tau)$. All other connections remain unaffected by $f$. One easily checks that $f(\tau)\in T_{N}$ and $f(f(\tau))=\tau$, hence the map is bijective. Since the root node has spin $\sigma_{1}=+1$ and the second node has the opposite spin $\sigma_{2}=-1$, property \eqref{eq:bijection} follows.
Let us denote all realizations that lead to a positive, negative or zero magnetization by $T_{N,+}, T_{N,-}, T_{N,0}$, respectively. Then the sets $T_{N,+}, T_{N,-}, T_{N,0}$ are pairwise disjoint and their union is
\begin{align}
	T_{N}= T_{N,+} \cup T_{N,-}\cup T_{N,0}.
	\label{eq:intersectiontreerealizations}
\end{align}
The expectation value of the magnetization of the RRT is
\begin{align}
	\langle M_{N}\rangle = \frac{1}{|T_{N}|}&\sum_{\tau\in T_{N}}M_{N}(\tau)
	\notag \\
	=\frac{1}{|T_{N}|}& \bigg( \sum_{\tau \in T_{N,+}} M_{N}(\tau) + \sum_{\tau \in T_{N,-}} M_{N}(\tau) 
	\notag \\
	&+ \sum_{\tau \in T_{N,0}} M_{N}(\tau)\bigg)
	\notag \\
	=\frac{1}{|T_{N}|}&\bigg( \sum_{\tau \in T_{N,+}} M_{N}(\tau) + \sum_{\tau \in T_{N,-}} M_{N}(\tau) \bigg)
	\notag \\
	=\frac{1}{|T_{N}|}&\bigg( \sum_{\tau \in T_{N,+}} M_{N}(\tau) + \sum_{\tau \in T_{N,+}} M_{N}(f(\tau)) \bigg)
	\notag \\
	=\frac{1}{|T_{N}|}& \sum_{\tau \in T_{N,+}} (M_{N}(\tau) + M_{N}(f(\tau))) = 0,
	\label{eq:magnetizationantialignedspins}
\end{align}
where we used that for all $\tau\in T_{N,0}$ $M_{N}(\tau)=0$ and that due to the bijectivity of $f$ and the reflection property \eqref{eq:bijection} we have $T_{N,-}=f(T_{N,+})$.
The considerations are of course equivalent if the root has spin $\sigma_{1}=-1$. Hence in this case the expectation value of the magnetization is minus one for $N=1$ and zero for $N\ge 2$.

\section{Second moment of number of nodes in first generation child clusters\label{app:b}}

We are calculating the expectation value of the number of nodes that are in child clusters of the first generation squared. Therefore we consider the LERW with the parameter $s=1-p-\varepsilon$.
We further modify the construction of the memory forest a little. Each time the elephant decides to go to sleep or to do the opposite of what it has done in the past the corresponding edge will be deleted.
In addition, given that the elephant decides to do the same as it did in the past, the corresponding edge will be deleted as well with probability $\varepsilon/p$. Thus each edge will be deleted with probability $1-p+\varepsilon$ and it will be kept with probability $p-\varepsilon$.
With this setup we find for the magnetization squared
\begin{align}
	&\langle x_{t=N}^{2}\rangle=\langle M_{N}^{2}\rangle
	\label{eq:magnetizationsquaredlazyepsilon}
	\\
	&= \langle |c_1|_{p-\varepsilon}^{2}\rangle + \frac{2\varepsilon}{1-p+\varepsilon} \sum_{i, c_{i} \text{of $1$st gen.}} \langle |c_{i}|_{p-\varepsilon}^{2}\rangle 
	\notag \\
	&\phantom{=}+ \frac{4 \varepsilon^{2}}{(1-p+\varepsilon)^{2}}\sum_{i, c_{i} \text{of $2$nd gen.}}\langle |c_{i}|_{p-\varepsilon}^{2}\rangle + \mathcal{O}(\varepsilon^{3})
	\notag \\
	&=\frac{1}{p-3\varepsilon}\left( \frac{\Gamma(2p-2\varepsilon+N)}{\Gamma(2p-2\varepsilon)\Gamma(N)} - \frac{\Gamma(p+\varepsilon+N)}{\Gamma(p+\varepsilon)\Gamma(N)} \right),
	\notag
\end{align}
where the lower index $_{p-\varepsilon}$ indicates that each edge was kept with probability $p-\varepsilon$. We used Eq.~\eqref{eq:secondmomentresultlazy} in the last line. When plugging Eq.~\eqref{eq:rootclustersecondmoment} in with the parameter $p-\varepsilon$, we can compare coefficients in different powers of $\varepsilon$. The terms of zeroth order in $\varepsilon$ just reproduce \eqref{eq:rootclustersecondmoment} for $\varepsilon\rightarrow 0$. Comparing terms linear in $\varepsilon$ we find
\begin{align}
	&\sum_{i, c_i \text{of $1$st gen.}} \langle |c_{i}|_{p}^{2}\rangle =
	\frac{1-p}{p}\bigg[ \frac{1}{p} \frac{\Gamma(2p+N)}{\Gamma(2p)\Gamma(N)} 
	\notag \\
	&+ \frac{\Gamma(p+N)}{\Gamma(p)\Gamma(N)}\Big(\Psi_{0}(p)- \Psi_{0}(p+N)-\frac{1}{p} \Big) \bigg]
	\label{eq:sumfirstgensquared}
\end{align}
where $\Psi_{k}$ denotes the polygamma function.

Replacing $p$ with $p-\varepsilon$ in Eq.~\eqref{eq:sumfirstgensquared} and plugging it in Eq.~\eqref{eq:magnetizationsquaredlazyepsilon} we can compare coefficients of $\varepsilon^{2}$ in Eq.~\eqref{eq:magnetizationsquaredlazyepsilon} to find
\begin{align}
	&\sum_{i, c_{i} \text{of $2$nd gen.}} \langle |c_{i}|_{p}^{2}\rangle = \frac{(1-p)^{2}}{p^{2}} \bigg[\frac{1}{p} \frac{\Gamma(2p+N)}{\Gamma(2p)\Gamma(N)} 
	+\frac{\Gamma(p+N)}{\Gamma(p)\Gamma(N)}
	\notag
	\\
	&\times\Big(-\frac{1}{p} + \Psi_{0}(p)  - \Psi_{0}(p+N) - \frac{p}{2}\big(\Psi_{0}^{2}(p) 
	+ \Psi_{0}^{2}(p+N) 
	\notag \\
	&-2\Psi_{0}(p)\Psi_{0}(p+N) - \Psi_{1}(p) + \Psi_{1}(p+N) \big) \Big) \bigg].
	\label{eq:sumsecondgensquared}
\end{align}
Now we consider the original memory forest of LERW. That means each time the elephant is doing the same as it did in the past, the corresponding edge is kept and otherwise it is deleted. That means it is kept with probability $p$ and deleted with probability $1-p$. We still consider the sleeping probability $s=1-p-\varepsilon$.
Then we find for the magnetization squared
\begin{align}
	\langle x_{t}^{2}\rangle&= \langle M_{N}^{2}\rangle 
	\notag \\
	&= \langle |c_{1}|^{2}\rangle + \frac{\varepsilon}{1-p} \sum_{i, c_{i} \text{of $1$st gen.}} \Big( \langle |c_{i}|^{2}\rangle - 2\langle |c_1||c_{i}|\rangle  \Big)
	\notag \\
	&\phantom{=} + \frac{\varepsilon^{2}}{(1-p)^{2}} \Big( \sum_{i<j, c_{i/j} \text{of $1$st gen.}} 2\langle |c_{i}||c_{j}|\rangle 
	\notag \\
	&\phantom{=} + \sum_{i, c_{i} \text{of $2$nd gen.}} \langle |c_{i}|^{2}\rangle \Big)
	\notag \\
	&=\frac{1}{p-3\varepsilon}\left( \frac{\Gamma(2p-2\varepsilon+N)}{\Gamma(2p-2\varepsilon)\Gamma(N)} - \frac{\Gamma(p+\varepsilon+N)}{\Gamma(p+\varepsilon)\Gamma(N)} \right).
	\label{eq:magnetizationsquaredlazyepsilon2}
\end{align}
Comparing terms of Eq.~\eqref{eq:magnetizationsquaredlazyepsilon2} that are linear in $\varepsilon$ we find 
\begin{align}
	&\sum_{i, c_{i} \text{of $1$st gen.}} \big( \langle |c_{i}|^{2}\rangle - 2\langle |c_{1}| |c_{i}|\rangle \big)
	\label{eq:correlationsfirstgen}
	\\
	&=\frac{1-p}{p}\bigg[ \frac{\Gamma(2p+N)}{\Gamma(2p)\Gamma(N)}\Big(\frac{3}{p} +2\Psi_{0}(2p)-2 \Psi_{0}(2p+N) \Big)
	\notag \\
	&\phantom{=} + \frac{\Gamma(p+N)}{\Gamma(p)\Gamma(N)} \Big(-\frac{3}{p} + \Psi_{0}(p) - \Psi_{0}(p+N) \Big)\bigg].
	\notag
\end{align}
From Eqs.~\eqref{eq:sumfirstgensquared} and \eqref{eq:correlationsfirstgen} we obtain the correlations between the root cluster size and the number of nodes in first generation child clusters \eqref{eq:correlationrootfirstgen}.

Comparing terms of Eq.~\eqref{eq:magnetizationsquaredlazyepsilon2} that are quadratic in $\varepsilon$ we find 
\begin{align}
	&\sum_{i<j, c_{i/j} \text{of $1$st gen.}} 2 \langle |c_{i}||c_{j}|\rangle + \sum_{i, c_{i} \text{of $2$nd gen.}} \langle |c_{i}|^{2}\rangle 
	\notag 
	\\
	&= \frac{(1-p)^{2}}{p^{2}}\bigg[ \frac{\Gamma(2p+N)}{\Gamma(2p)\Gamma(N)}\Big(\frac{9}{p} + 6 \Psi_{0}(2p) - 6 \Psi_{0}(2p+N)
	\notag 
	\\
	&\phantom{=}+ 2p\big(\Psi_{0}^{2}(2p)+\Psi_{0}^{2}(2p+N) -2\Psi_{0}(2p)\Psi_{0}(2p+N) 
	\notag
	\\
	&\phantom{=}- \Psi_{1}(2p) + \Psi_{1}(2p+N) \big)  \Big)
	\notag
	\\
	&\phantom{=} + \frac{\Gamma(p+N)}{\Gamma(p)\Gamma(N)} \Big(-\frac{9}{p} + 3 \Psi_{0}(p) - 3 \Psi_{0}(p+N) 
	\notag
	\\
	&\phantom{=} -\frac{p}{2} \big(\Psi_{0}^{2}(p) + \Psi_{0}^{2}(p+N) + 2\Psi_{0}(p)\Psi_{0}(p+N) 
	\notag
	\\
	&\phantom{=}- \Psi_{1}(p) + \Psi_{1}(p+N) \big) \Big) \bigg].
	\label{eq:secondgenmixedterm}
\end{align}
Subtracting Eq.~\eqref{eq:sumsecondgensquared} from and adding Eq.~\eqref{eq:sumfirstgensquared} to Eq.~\eqref{eq:secondgenmixedterm}, we obtain the expectation value of the number of nodes in first generation child clusters squared. The result is given in Eq.~\eqref{eq:firstgensecondmoment}. One can continue to calculate correlations and the second moment of the number of nodes in $k$th generation child clusters, evaluating terms of higher order in $\varepsilon$.
\end{appendix}

\begin{acknowledgments}
	The author thanks the International Max Planck Research School, Mathematics in the Sciences, Leipzig for supporting part of this work with a scholarship. The author thanks Ulrich Behn and Marco Müller for valuable remarks on the manuscript.
\end{acknowledgments}

\bibliography{literatur.bib}

\begin{thebibliography}{33}%
\makeatletter
\providecommand \@ifxundefined [1]{%
 \@ifx{#1\undefined}
}%
\providecommand \@ifnum [1]{%
 \ifnum #1\expandafter \@firstoftwo
 \else \expandafter \@secondoftwo
 \fi
}%
\providecommand \@ifx [1]{%
 \ifx #1\expandafter \@firstoftwo
 \else \expandafter \@secondoftwo
 \fi
}%
\providecommand \natexlab [1]{#1}%
\providecommand \enquote  [1]{``#1''}%
\providecommand \bibnamefont  [1]{#1}%
\providecommand \bibfnamefont [1]{#1}%
\providecommand \citenamefont [1]{#1}%
\providecommand \href@noop [0]{\@secondoftwo}%
\providecommand \href [0]{\begingroup \@sanitize@url \@href}%
\providecommand \@href[1]{\@@startlink{#1}\@@href}%
\providecommand \@@href[1]{\endgroup#1\@@endlink}%
\providecommand \@sanitize@url [0]{\catcode `\\12\catcode `\$12\catcode
  `\&12\catcode `\#12\catcode `\^12\catcode `\_12\catcode `\%12\relax}%
\providecommand \@@startlink[1]{}%
\providecommand \@@endlink[0]{}%
\providecommand \url  [0]{\begingroup\@sanitize@url \@url }%
\providecommand \@url [1]{\endgroup\@href {#1}{\urlprefix }}%
\providecommand \urlprefix  [0]{URL }%
\providecommand \Eprint [0]{\href }%
\providecommand \doibase [0]{http://dx.doi.org/}%
\providecommand \selectlanguage [0]{\@gobble}%
\providecommand \bibinfo  [0]{\@secondoftwo}%
\providecommand \bibfield  [0]{\@secondoftwo}%
\providecommand \translation [1]{[#1]}%
\providecommand \BibitemOpen [0]{}%
\providecommand \bibitemStop [0]{}%
\providecommand \bibitemNoStop [0]{.\EOS\space}%
\providecommand \EOS [0]{\spacefactor3000\relax}%
\providecommand \BibitemShut  [1]{\csname bibitem#1\endcsname}%
\let\auto@bib@innerbib\@empty
\bibitem [{\citenamefont {Brockmann}\ \emph {et~al.}(2006)\citenamefont
  {Brockmann}, \citenamefont {Hufnagel},\ and\ \citenamefont {Geisel}}]{BHG06}%
  \BibitemOpen
  \bibfield  {author} {\bibinfo {author} {\bibfnamefont {D.}~\bibnamefont
  {Brockmann}}, \bibinfo {author} {\bibfnamefont {L.}~\bibnamefont {Hufnagel}},
  \ and\ \bibinfo {author} {\bibfnamefont {T.}~\bibnamefont {Geisel}},\
  }\href@noop {} {\bibfield  {journal} {\bibinfo  {journal} {Nature}\ }\textbf
  {\bibinfo {volume} {439}},\ \bibinfo {pages} {462} (\bibinfo {year}
  {2006})}\BibitemShut {NoStop}%
\bibitem [{\citenamefont {Weiss}\ \emph {et~al.}(2004)\citenamefont {Weiss},
  \citenamefont {Elsner}, \citenamefont {Kartberg},\ and\ \citenamefont
  {Nilsson}}]{WEKT04}%
  \BibitemOpen
  \bibfield  {author} {\bibinfo {author} {\bibfnamefont {M.}~\bibnamefont
  {Weiss}}, \bibinfo {author} {\bibfnamefont {M.}~\bibnamefont {Elsner}},
  \bibinfo {author} {\bibfnamefont {F.}~\bibnamefont {Kartberg}}, \ and\
  \bibinfo {author} {\bibfnamefont {T.}~\bibnamefont {Nilsson}},\ }\href@noop
  {} {\bibfield  {journal} {\bibinfo  {journal} {Biophys. J.}\ }\textbf
  {\bibinfo {volume} {87}},\ \bibinfo {pages} {3518} (\bibinfo {year}
  {2004})}\BibitemShut {NoStop}%
\bibitem [{\citenamefont {Sch{\"u}tz}\ \emph {et~al.}(1997)\citenamefont
  {Sch{\"u}tz}, \citenamefont {Schindler},\ and\ \citenamefont
  {Schmidt}}]{SSS97}%
  \BibitemOpen
  \bibfield  {author} {\bibinfo {author} {\bibfnamefont {G.}~\bibnamefont
  {Sch{\"u}tz}}, \bibinfo {author} {\bibfnamefont {H.}~\bibnamefont
  {Schindler}}, \ and\ \bibinfo {author} {\bibfnamefont {T.}~\bibnamefont
  {Schmidt}},\ }\href@noop {} {\bibfield  {journal} {\bibinfo  {journal}
  {Biophys. J.}\ }\textbf {\bibinfo {volume} {73}},\ \bibinfo {pages} {1073}
  (\bibinfo {year} {1997})}\BibitemShut {NoStop}%
\bibitem [{\citenamefont {Koch}\ and\ \citenamefont {Brady}(1988)}]{KB88}%
  \BibitemOpen
  \bibfield  {author} {\bibinfo {author} {\bibfnamefont {D.~L.}\ \bibnamefont
  {Koch}}\ and\ \bibinfo {author} {\bibfnamefont {J.~F.}\ \bibnamefont
  {Brady}},\ }\href@noop {} {\bibfield  {journal} {\bibinfo  {journal} {Phys.
  Fluids}\ }\textbf {\bibinfo {volume} {31}},\ \bibinfo {pages} {965} (\bibinfo
  {year} {1988})}\BibitemShut {NoStop}%
\bibitem [{\citenamefont {Wong}\ \emph {et~al.}(2004)\citenamefont {Wong},
  \citenamefont {Gardel}, \citenamefont {Reichman}, \citenamefont {Weeks},
  \citenamefont {Valentine}, \citenamefont {Bausch},\ and\ \citenamefont
  {Weitz}}]{anomalactin04}%
  \BibitemOpen
  \bibfield  {author} {\bibinfo {author} {\bibfnamefont {I.~Y.}\ \bibnamefont
  {Wong}}, \bibinfo {author} {\bibfnamefont {M.~L.}\ \bibnamefont {Gardel}},
  \bibinfo {author} {\bibfnamefont {D.~R.}\ \bibnamefont {Reichman}}, \bibinfo
  {author} {\bibfnamefont {E.~R.}\ \bibnamefont {Weeks}}, \bibinfo {author}
  {\bibfnamefont {M.~T.}\ \bibnamefont {Valentine}}, \bibinfo {author}
  {\bibfnamefont {A.~R.}\ \bibnamefont {Bausch}}, \ and\ \bibinfo {author}
  {\bibfnamefont {D.~A.}\ \bibnamefont {Weitz}},\ }\href {\doibase
  10.1103/PhysRevLett.92.178101} {\bibfield  {journal} {\bibinfo  {journal}
  {Phys. Rev. Lett.}\ }\textbf {\bibinfo {volume} {92}},\ \bibinfo {pages}
  {178101} (\bibinfo {year} {2004})}\BibitemShut {NoStop}%
\bibitem [{\citenamefont {Mandelbrot}\ and\ \citenamefont
  {Van~Ness}(1968)}]{MN68}%
  \BibitemOpen
  \bibfield  {author} {\bibinfo {author} {\bibfnamefont {B.~B.}\ \bibnamefont
  {Mandelbrot}}\ and\ \bibinfo {author} {\bibfnamefont {J.~W.}\ \bibnamefont
  {Van~Ness}},\ }\href@noop {} {\bibfield  {journal} {\bibinfo  {journal} {SIAM
  Rev.}\ }\textbf {\bibinfo {volume} {10}},\ \bibinfo {pages} {422} (\bibinfo
  {year} {1968})}\BibitemShut {NoStop}%
\bibitem [{\citenamefont {Gefen}\ \emph {et~al.}(1983)\citenamefont {Gefen},
  \citenamefont {Aharony},\ and\ \citenamefont {Alexander}}]{GAA83}%
  \BibitemOpen
  \bibfield  {author} {\bibinfo {author} {\bibfnamefont {Y.}~\bibnamefont
  {Gefen}}, \bibinfo {author} {\bibfnamefont {A.}~\bibnamefont {Aharony}}, \
  and\ \bibinfo {author} {\bibfnamefont {S.}~\bibnamefont {Alexander}},\ }\href
  {\doibase 10.1103/PhysRevLett.50.77} {\bibfield  {journal} {\bibinfo
  {journal} {Phys. Rev. Lett.}\ }\textbf {\bibinfo {volume} {50}},\ \bibinfo
  {pages} {77} (\bibinfo {year} {1983})}\BibitemShut {NoStop}%
\bibitem [{\citenamefont {Scher}\ and\ \citenamefont {Montroll}(1975)}]{SM75}%
  \BibitemOpen
  \bibfield  {author} {\bibinfo {author} {\bibfnamefont {H.}~\bibnamefont
  {Scher}}\ and\ \bibinfo {author} {\bibfnamefont {E.~W.}\ \bibnamefont
  {Montroll}},\ }\href {\doibase 10.1103/PhysRevB.12.2455} {\bibfield
  {journal} {\bibinfo  {journal} {Phys. Rev. B}\ }\textbf {\bibinfo {volume}
  {12}},\ \bibinfo {pages} {2455} (\bibinfo {year} {1975})}\BibitemShut
  {NoStop}%
\bibitem [{\citenamefont {Metzler}\ and\ \citenamefont {Klafter}(2000)}]{MK00}%
  \BibitemOpen
  \bibfield  {author} {\bibinfo {author} {\bibfnamefont {R.}~\bibnamefont
  {Metzler}}\ and\ \bibinfo {author} {\bibfnamefont {J.}~\bibnamefont
  {Klafter}},\ }\href {\doibase
  http://dx.doi.org/10.1016/S0370-1573(00)00070-3} {\bibfield  {journal}
  {\bibinfo  {journal} {Phys. Rep.}\ }\textbf {\bibinfo {volume} {339}},\
  \bibinfo {pages} {1 } (\bibinfo {year} {2000})}\BibitemShut {NoStop}%
\bibitem [{\citenamefont {Metzler}\ and\ \citenamefont {Klafter}(2004)}]{MK04}%
  \BibitemOpen
  \bibfield  {author} {\bibinfo {author} {\bibfnamefont {R.}~\bibnamefont
  {Metzler}}\ and\ \bibinfo {author} {\bibfnamefont {J.}~\bibnamefont
  {Klafter}},\ }\href {http://stacks.iop.org/0305-4470/37/i=31/a=R01}
  {\bibfield  {journal} {\bibinfo  {journal} {J. Phys. A - Math. Gen.}\
  }\textbf {\bibinfo {volume} {37}},\ \bibinfo {pages} {R161} (\bibinfo {year}
  {2004})}\BibitemShut {NoStop}%
\bibitem [{\citenamefont {Sch\"utz}\ and\ \citenamefont
  {Trimper}(2004)}]{ST04}%
  \BibitemOpen
  \bibfield  {author} {\bibinfo {author} {\bibfnamefont {G.~M.}\ \bibnamefont
  {Sch\"utz}}\ and\ \bibinfo {author} {\bibfnamefont {S.}~\bibnamefont
  {Trimper}},\ }\href {\doibase 10.1103/PhysRevE.70.045101} {\bibfield
  {journal} {\bibinfo  {journal} {Phys. Rev. E}\ }\textbf {\bibinfo {volume}
  {70}},\ \bibinfo {pages} {045101} (\bibinfo {year} {2004})}\BibitemShut
  {NoStop}%
\bibitem [{\citenamefont {Cressoni}\ \emph {et~al.}(2007)\citenamefont
  {Cressoni}, \citenamefont {da~Silva},\ and\ \citenamefont
  {Viswanathan}}]{CSV07}%
  \BibitemOpen
  \bibfield  {author} {\bibinfo {author} {\bibfnamefont {J.~C.}\ \bibnamefont
  {Cressoni}}, \bibinfo {author} {\bibfnamefont {M.~A.~A.}\ \bibnamefont
  {da~Silva}}, \ and\ \bibinfo {author} {\bibfnamefont {G.~M.}\ \bibnamefont
  {Viswanathan}},\ }\href {\doibase 10.1103/PhysRevLett.98.070603} {\bibfield
  {journal} {\bibinfo  {journal} {Phys. Rev. Lett.}\ }\textbf {\bibinfo
  {volume} {98}},\ \bibinfo {pages} {070603} (\bibinfo {year}
  {2007})}\BibitemShut {NoStop}%
\bibitem [{\citenamefont {Kumar}\ \emph {et~al.}(2010)\citenamefont {Kumar},
  \citenamefont {Harbola},\ and\ \citenamefont {Lindenberg}}]{KHL10}%
  \BibitemOpen
  \bibfield  {author} {\bibinfo {author} {\bibfnamefont {N.}~\bibnamefont
  {Kumar}}, \bibinfo {author} {\bibfnamefont {U.}~\bibnamefont {Harbola}}, \
  and\ \bibinfo {author} {\bibfnamefont {K.}~\bibnamefont {Lindenberg}},\
  }\href {\doibase 10.1103/PhysRevE.82.021101} {\bibfield  {journal} {\bibinfo
  {journal} {Phys. Rev. E}\ }\textbf {\bibinfo {volume} {82}},\ \bibinfo
  {pages} {021101} (\bibinfo {year} {2010})}\BibitemShut {NoStop}%
\bibitem [{\citenamefont {Cressoni}\ \emph {et~al.}(2012)\citenamefont
  {Cressoni}, \citenamefont {Viswanathan}, \citenamefont {Ferreira},\ and\
  \citenamefont {da~Silva}}]{CVFS12}%
  \BibitemOpen
  \bibfield  {author} {\bibinfo {author} {\bibfnamefont {J.~C.}\ \bibnamefont
  {Cressoni}}, \bibinfo {author} {\bibfnamefont {G.~M.}\ \bibnamefont
  {Viswanathan}}, \bibinfo {author} {\bibfnamefont {A.~S.}\ \bibnamefont
  {Ferreira}}, \ and\ \bibinfo {author} {\bibfnamefont {M.~A.~A.}\ \bibnamefont
  {da~Silva}},\ }\href {\doibase 10.1103/PhysRevE.86.022103} {\bibfield
  {journal} {\bibinfo  {journal} {Phys. Rev. E}\ }\textbf {\bibinfo {volume}
  {86}},\ \bibinfo {pages} {022103} (\bibinfo {year} {2012})}\BibitemShut
  {NoStop}%
\bibitem [{\citenamefont {Cressoni}\ \emph {et~al.}(2013)\citenamefont
  {Cressoni}, \citenamefont {Viswanathan},\ and\ \citenamefont
  {da~Silva}}]{CVS13}%
  \BibitemOpen
  \bibfield  {author} {\bibinfo {author} {\bibfnamefont {J.~C.}\ \bibnamefont
  {Cressoni}}, \bibinfo {author} {\bibfnamefont {G.~M.}\ \bibnamefont
  {Viswanathan}}, \ and\ \bibinfo {author} {\bibfnamefont {M.~A.~A.}\
  \bibnamefont {da~Silva}},\ }\href
  {http://stacks.iop.org/1751-8121/46/i=50/a=505002} {\bibfield  {journal}
  {\bibinfo  {journal} {J. Phys. A - Math. Theor.}\ }\textbf {\bibinfo {volume}
  {46}},\ \bibinfo {pages} {505002} (\bibinfo {year} {2013})}\BibitemShut
  {NoStop}%
\bibitem [{\citenamefont {Harbola}\ \emph {et~al.}(2014)\citenamefont
  {Harbola}, \citenamefont {Kumar},\ and\ \citenamefont {Lindenberg}}]{HKL14}%
  \BibitemOpen
  \bibfield  {author} {\bibinfo {author} {\bibfnamefont {U.}~\bibnamefont
  {Harbola}}, \bibinfo {author} {\bibfnamefont {N.}~\bibnamefont {Kumar}}, \
  and\ \bibinfo {author} {\bibfnamefont {K.}~\bibnamefont {Lindenberg}},\
  }\href {\doibase 10.1103/PhysRevE.90.022136} {\bibfield  {journal} {\bibinfo
  {journal} {Phys. Rev. E}\ }\textbf {\bibinfo {volume} {90}},\ \bibinfo
  {pages} {022136} (\bibinfo {year} {2014})}\BibitemShut {NoStop}%
\bibitem [{\citenamefont {Kim}(2014)}]{Kim14}%
  \BibitemOpen
  \bibfield  {author} {\bibinfo {author} {\bibfnamefont {H.-J.}\ \bibnamefont
  {Kim}},\ }\href {\doibase 10.1103/PhysRevE.90.012103} {\bibfield  {journal}
  {\bibinfo  {journal} {Phys. Rev. E}\ }\textbf {\bibinfo {volume} {90}},\
  \bibinfo {pages} {012103} (\bibinfo {year} {2014})}\BibitemShut {NoStop}%
\bibitem [{\citenamefont {da~Silva}\ \emph {et~al.}(2015)\citenamefont
  {da~Silva}, \citenamefont {Viswanathan},\ and\ \citenamefont
  {Cressoni}}]{SVC15}%
  \BibitemOpen
  \bibfield  {author} {\bibinfo {author} {\bibfnamefont {M.}~\bibnamefont
  {da~Silva}}, \bibinfo {author} {\bibfnamefont {G.}~\bibnamefont
  {Viswanathan}}, \ and\ \bibinfo {author} {\bibfnamefont {J.}~\bibnamefont
  {Cressoni}},\ }\href {\doibase http://dx.doi.org/10.1016/j.physa.2014.11.047}
  {\bibfield  {journal} {\bibinfo  {journal} {Physica A}\ }\textbf {\bibinfo
  {volume} {421}},\ \bibinfo {pages} {522 } (\bibinfo {year}
  {2015})}\BibitemShut {NoStop}%
\bibitem [{\citenamefont {Yule}(1925)}]{Yule25}%
  \BibitemOpen
  \bibfield  {author} {\bibinfo {author} {\bibfnamefont {G.~U.}\ \bibnamefont
  {Yule}},\ }\href {http://www.jstor.org/stable/92117} {\bibfield  {journal}
  {\bibinfo  {journal} {Philos. T. R. Soc. Lon. B}\ }\textbf {\bibinfo {volume}
  {213}},\ \bibinfo {pages} {pp. 21} (\bibinfo {year} {1925})}\BibitemShut
  {NoStop}%
\bibitem [{\citenamefont {Simon}(1955)}]{Simon55}%
  \BibitemOpen
  \bibfield  {author} {\bibinfo {author} {\bibfnamefont {H.~A.}\ \bibnamefont
  {Simon}},\ }\href@noop {} {\bibfield  {journal} {\bibinfo  {journal}
  {Biometrika}\ }\textbf {\bibinfo {volume} {42}},\ \bibinfo {pages} {425}
  (\bibinfo {year} {1955})}\BibitemShut {NoStop}%
\bibitem [{\citenamefont {Price}(1965)}]{Price65}%
  \BibitemOpen
  \bibfield  {author} {\bibinfo {author} {\bibfnamefont {D.}~\bibnamefont
  {Price}},\ }\href@noop {} {\bibfield  {journal} {\bibinfo  {journal}
  {Science}\ }\textbf {\bibinfo {volume} {149}},\ \bibinfo {pages} {157}
  (\bibinfo {year} {1965})}\BibitemShut {NoStop}%
\bibitem [{\citenamefont {Barab{\'a}si}\ and\ \citenamefont
  {Albert}(1999)}]{BA99}%
  \BibitemOpen
  \bibfield  {author} {\bibinfo {author} {\bibfnamefont {A.-L.}\ \bibnamefont
  {Barab{\'a}si}}\ and\ \bibinfo {author} {\bibfnamefont {R.}~\bibnamefont
  {Albert}},\ }\href@noop {} {\bibfield  {journal} {\bibinfo  {journal}
  {Science}\ }\textbf {\bibinfo {volume} {286}},\ \bibinfo {pages} {509}
  (\bibinfo {year} {1999})}\BibitemShut {NoStop}%
\bibitem [{\citenamefont {Bertoin}(2014)}]{Bertoin14}%
  \BibitemOpen
  \bibfield  {author} {\bibinfo {author} {\bibfnamefont {J.}~\bibnamefont
  {Bertoin}},\ }\href@noop {} {\bibfield  {journal} {\bibinfo  {journal}
  {Random Struct. Algor.}\ }\textbf {\bibinfo {volume} {44}},\ \bibinfo {pages}
  {29} (\bibinfo {year} {2014})}\BibitemShut {NoStop}%
\bibitem [{\citenamefont {{Baur}}(2014)}]{Baur14}%
  \BibitemOpen
  \bibfield  {author} {\bibinfo {author} {\bibfnamefont {E.}~\bibnamefont
  {{Baur}}},\ }\href@noop {} {\bibfield  {journal} {\bibinfo  {journal} {ArXiv
  e-prints}\ } (\bibinfo {year} {2014})},\ \Eprint
  {http://arxiv.org/abs/1407.2508} {arXiv:1407.2508 [math.PR]} \BibitemShut
  {NoStop}%
\bibitem [{\citenamefont {Baur}\ and\ \citenamefont {Bertoin}(2014)}]{BB14}%
  \BibitemOpen
  \bibfield  {author} {\bibinfo {author} {\bibfnamefont {E.}~\bibnamefont
  {Baur}}\ and\ \bibinfo {author} {\bibfnamefont {J.}~\bibnamefont {Bertoin}},\
  }in\ \href {\doibase 10.1007/978-3-319-11292-3_3} {\emph {\bibinfo
  {booktitle} {Stochastic Analysis and Applications 2014}}},\ \bibinfo {series}
  {Springer Proceedings in Mathematics \& Statistics}, Vol.\ \bibinfo {volume}
  {100},\ \bibinfo {editor} {edited by\ \bibinfo {editor} {\bibfnamefont
  {D.}~\bibnamefont {Crisan}}, \bibinfo {editor} {\bibfnamefont
  {B.}~\bibnamefont {Hambly}}, \ and\ \bibinfo {editor} {\bibfnamefont
  {T.}~\bibnamefont {Zariphopoulou}}}\ (\bibinfo  {publisher} {Springer
  International Publishing},\ \bibinfo {year} {2014})\ pp.\ \bibinfo {pages}
  {51--76}\BibitemShut {NoStop}%
\bibitem [{\citenamefont {Meir}\ and\ \citenamefont {Moon}(1974)}]{MM74}%
  \BibitemOpen
  \bibfield  {author} {\bibinfo {author} {\bibfnamefont {A.}~\bibnamefont
  {Meir}}\ and\ \bibinfo {author} {\bibfnamefont {J.}~\bibnamefont {Moon}},\
  }\href@noop {} {\bibfield  {journal} {\bibinfo  {journal} {Math. Biosci.}\
  }\textbf {\bibinfo {volume} {21}},\ \bibinfo {pages} {173} (\bibinfo {year}
  {1974})}\BibitemShut {NoStop}%
\bibitem [{\citenamefont {Iksanov}\ and\ \citenamefont
  {M{\"o}hle}(2007)}]{IM07}%
  \BibitemOpen
  \bibfield  {author} {\bibinfo {author} {\bibfnamefont {A.}~\bibnamefont
  {Iksanov}}\ and\ \bibinfo {author} {\bibfnamefont {M.}~\bibnamefont
  {M{\"o}hle}},\ }\href@noop {} {\bibfield  {journal} {\bibinfo  {journal}
  {Electron. Commun. Prob.}\ }\textbf {\bibinfo {volume} {12}},\ \bibinfo
  {pages} {28} (\bibinfo {year} {2007})}\BibitemShut {NoStop}%
\bibitem [{\citenamefont {Kuba}\ and\ \citenamefont {Panholzer}(2014)}]{KP14}%
  \BibitemOpen
  \bibfield  {author} {\bibinfo {author} {\bibfnamefont {M.}~\bibnamefont
  {Kuba}}\ and\ \bibinfo {author} {\bibfnamefont {A.}~\bibnamefont
  {Panholzer}},\ }\href@noop {} {\bibfield  {journal} {\bibinfo  {journal}
  {Online J. Anal. Comb.}\ }\textbf {\bibinfo {volume} {9}},\ \bibinfo {pages}
  {26} (\bibinfo {year} {2014})}\BibitemShut {NoStop}%
\bibitem [{\citenamefont {Marzouk}(2016)}]{Marzouk14}%
  \BibitemOpen
  \bibfield  {author} {\bibinfo {author} {\bibfnamefont {C.}~\bibnamefont
  {Marzouk}},\ }\href {\doibase http://dx.doi.org/10.1016/j.spa.2015.08.006}
  {\bibfield  {journal} {\bibinfo  {journal} {Stoch. Proc. Appl.}\ }\textbf
  {\bibinfo {volume} {126}},\ \bibinfo {pages} {265 } (\bibinfo {year}
  {2016})}\BibitemShut {NoStop}%
\bibitem [{\citenamefont {{Baur}}\ and\ \citenamefont
  {{Bertoin}}(2015)}]{BB15}%
  \BibitemOpen
  \bibfield  {author} {\bibinfo {author} {\bibfnamefont {E.}~\bibnamefont
  {{Baur}}}\ and\ \bibinfo {author} {\bibfnamefont {J.}~\bibnamefont
  {{Bertoin}}},\ }\href@noop {} {\bibfield  {journal} {\bibinfo  {journal}
  {Electron. J. Probab.}\ }\textbf {\bibinfo {volume} {20}},\ \bibinfo {pages}
  {1} (\bibinfo {year} {2015})}\BibitemShut {NoStop}%
\bibitem [{\citenamefont {Kalay}\ and\ \citenamefont {Ben-Naim}(2015)}]{KB15}%
  \BibitemOpen
  \bibfield  {author} {\bibinfo {author} {\bibfnamefont {Z.}~\bibnamefont
  {Kalay}}\ and\ \bibinfo {author} {\bibfnamefont {E.}~\bibnamefont
  {Ben-Naim}},\ }\href@noop {} {\bibfield  {journal} {\bibinfo  {journal} {J.
  Phys. A - Math. Theor.}\ }\textbf {\bibinfo {volume} {48}},\ \bibinfo {pages}
  {045001} (\bibinfo {year} {2015})}\BibitemShut {NoStop}%
\bibitem [{\citenamefont {{K{\"u}rsten}}(2015)}]{Kuersten15}%
  \BibitemOpen
  \bibfield  {author} {\bibinfo {author} {\bibfnamefont {R.}~\bibnamefont
  {{K{\"u}rsten}}},\ }\href@noop {} {\bibfield  {journal} {\bibinfo  {journal}
  {ArXiv e-prints}\ } (\bibinfo {year} {2015})},\ \Eprint
  {http://arxiv.org/abs/1503.03302} {arXiv:1503.03302 [physics.data-an]}
  \BibitemShut {NoStop}%
\bibitem [{\citenamefont {Grimmett}(1999)}]{Grimmett99}%
  \BibitemOpen
  \bibfield  {author} {\bibinfo {author} {\bibfnamefont {G.~R.}\ \bibnamefont
  {Grimmett}},\ }\href {\doibase 10.1007/978-3-662-03981-6} {\emph {\bibinfo
  {title} {Percolation}}},\ \bibinfo {edition} {2nd}\ ed.\ (\bibinfo
  {publisher} {Springer Verlag},\ \bibinfo {address} {Berlin Heidelberg},\
  \bibinfo {year} {1999})\BibitemShut {NoStop}%
\end{thebibliography}%

\end{document}